# DC Voltage Control Dominated Oscillations in Electromechanical Frequency Range — Insights from Synchronous Coefficient Reduction Effect of IBR Plants and Network Interactions

Qiang Fu, *Member, IEEE*, Siqi Bu, *Senior Member, IEEE*

***Abstract*—Large-scale integration of inverter-based resources (IBRs) is widely known to cause subsynchronous oscillations. This paper investigates a distinct class of oscillations that occur in the electromechanical frequency range (EFR), referred to as EFOs, and are dominated by the DC voltage control (DVC) loops of grid-following IBRs rather than synchronous generator-like controls or phase-locked loops (PLLs). Although open-loop oscillation modes of DVC loops are subsynchronous, the corresponding closed-loop dominant oscillation mode related to DVC can shift downward into the EFR without retuning the control parameters. This frequency shift is caused by network-mediated interactions among the IBRs, which introduce a negative synchronous coefficient into the DVC oscillation loop. This synchronous coefficient reduction effect (SCRE) becomes stronger as the IBR plant scale increases and/or as the minimum eigenvalue of the network admittance matrix decreases. Furthermore, a complex vector-based stability analysis method is proposed, and a stability criterion for DVC-EFOs is derived. Finally, the analytical conclusions are validated using a practical power system configuration implemented in Simulink. The SCRE differs from traditional explanations, in which the open-loop oscillation modes are already located in the EFR, and provides new insight into DVC-dominated oscillations in the EFR for integration of IBR plants.**

## I. INTRODUCTION

By the end of 2025, global installed renewable power capacity had reached 5149 GW, with 692 GW added in 2025 alone [1]. Much of this growth is converter interfaced, such as solar PV, wind power, and battery energy storage systems. To enhance both capacity and operational efficiency, many inverter-based resources (IBRs) are aggregated into plants. The integration of IBRs changes the characteristics of AC power systems and may introduce wideband oscillation risks [2]. Therefore, these risks must be examined to ensure stability.

Most existing studies on grid-following (GFL) IBR-induced oscillations focus on frequencies above the electromechanical frequency range (EFR), especially subsynchronous oscillations (SSOs) [3]. In this range, phase-locked loop (PLL) dynamics can destabilize synchronization under weak grid connections [4], and the inner current control loop can interact with converter filters and grid impedance, thereby reducing the damping margin [5]. DC voltage control (DVC) has also been studied mainly from the perspectives of DC voltage stability and stability in the DC voltage time scale. The reported mechanisms include the negative impedance of constant-power DC loads [6], strong dynamic interactions between the DVC and PLL under weak grid connections [7], and open-loop modal resonance with other control loops [8]. However, existing studies generally locate DVC-dominated oscillations in the subsynchronous frequency range, which is consistent with the intrinsic time scale of the DVC loops. Whether a DVC-dominated oscillation can shift into the EFR without manual controller parameter tuning has not yet been comprehensively reported.

TABLE I
Different types of EFOs

| Type | Cause | Formation Mechanism |
|---|---|---|
| Traditional EFOs | Synchronous generators (SGs) | Negative damping of rotor dynamics [9]-[11] |
| | GFM-IBRs | Negative damping of SG-like control loop [12]-[13] |
| | GFL-IBRs | Synchronous instability of PLL, and manual DVC parameters tuning [14]-[15] |
| | Plant-level control | Plant-level control settings and delay [16]-[17] |
| DVC-EFOs | DVC loops of IBRs | Network-mediated interactions among IBRs, without manual control parameter tuning |

Regarding oscillations that occur in or close to the EFR, referred to as EFOs, the typical scenarios can be classified into four cases, as summarized in Table I. The first case is power systems involving SGs. In this case, IBRs do not usually provide new electromechanical oscillation modes, but affect existing SG electromechanical oscillations by changing operation states. As shown in [9], increasing the IBR penetration can decrease the damping of critical electromechanical oscillation modes. The Ontario system in [10] and the U.S. Eastern Interconnection in [11] further confirm that high IBR penetration may affect inter-area electromechanical oscillation modes.

The second case is IBRs employing controls whose inherent

Qiang Fu and Siqi Bu are with the Department of Electrical and Electronic Engineering, The Hong Kong Polytechnic University. This work is supported by the National Natural Science Foundation of China (Grant No. U25B20205), (Grant No. 52407129). Corresponding author: Siqi Bu, siqi.bu@gmail.com.

dynamics (can be evaluated by the open-loop oscillation modes of IBRs) are in or close to the EFR. The virtual synchronous generator (VSG) control is the most typical example because it introduces SG-like swing dynamics. In [12] and [13], the impacts of VSG control parameters on the EFOs were analyzed, and stable parameter regions were obtained [12], showing that VSG-IBRs can directly participate in EFOs [13]. For more generalized grid-forming (GFM) control, the dynamic modeling and oscillation issues of power electronics-based energy storage systems are summarized in [18].

The third case is the frequency shift of GFL-IBRs such that closed-loop oscillation modes are located in or close to the EFR. In [15], type-4 wind farms under weak-grid conditions were shown to reproduce 4-Hz EFOs, and the PLL was identified as a key factor affecting the oscillation frequency. Moreover, the manual tuning of controller parameters is a common way for reducing the open-loop and closed-loop oscillation frequencies. For example, tuning the DVC control parameters can shift the open-loop oscillation mode to the EFR [14]. However, this is rarely acceptable because the control parameters of GFL-IBRs should be selected to ensure a fast dynamic response such that the inherent DVC dynamics remain above the EFR.

The fourth case is associated with plant-level control delay. This mechanism is different from EFOs caused by SG dynamics and converter-level control loops. In [16], 0.1-Hz voltage and reactive power oscillations in solar PV plants were reported, with the oscillations appearing primarily in voltage and reactive power rather than in active power. Similar voltage oscillations have also been studied in renewable energy complexes in India [17]. In these scenarios, plant-level control settings, volt-var feedback, and communication delay between the plant-level controller and inverter-level control were considered major causes.

TABLE II
Differences between DVC-EFOs and Traditional EFO Mechanisms

| Type | Cause | No Plant-Level Control | No Inherent EFR Mode | No Controller Tuning | Not PLL Sync. Instability |
|---|---|---|---|---|---|
| Traditional EFOs | SGs [9]-[11] | √ | X | √ | √ |
| | SG-like control [12]-[13] | √ | X | √ | √ |
| | PLL [15] | √ | √ | √ | X |
| | DVC loop [14] | √ | X | X | √ |
| | Plant-level control [16]-[17] | X | √ | √ | √ |
| DVC-EFOs | DVC loops | √ | √ | √ | √ |

However, as summarized in Table II, the DVC-EFOs studied in this paper are fundamentally different from the above EFO mechanisms. First, they originate from converter-level DVC loops of GFL-IBRs rather than plant-level control. Therefore, explanations based on plant-level control delay are not applicable. Second, their dynamic source differs from that of SGs and SG-like IBRs. SGs and SG-like IBRs can inherently produce EFR modes, whereas the GFL-IBRs considered here do not inherently provide EFR modes, in particular, the open-loop DVC oscillation mode is inherently outside the EFR. Third, although EFOs associated with GFL-IBRs have been reported, they are typically attributed to PLL synchronization instability under weak grid connections or to DVC controller tuning that shifts the oscillation mode into the EFR. In contrast, the DVC-EFOs studied here occur without controller tuning and are not caused by PLL synchronization instability. To clarify these differences, Table II summarizes the features of DVC-EFOs compared with traditional EFO mechanisms. The contributions of this study are as follows.

1) Motivated by EFOs observed in a practical power system integrated with PV plants, this study provides new insight into their formation mechanism: EFOs can be induced by the DVC loops of GFL-IBRs even when the inherent DVC dynamics are in the subsynchronous frequency range. This differs from traditional EFOs caused by SGs and IBRs with SG-like controls, whose inherent dynamics are located in the EFR, as well as from those caused by GFL-IBRs by tuning control parameters.

2) The synchronous coefficient reduction effect (SCRE) is revealed, explaining why an open-loop subsynchronous oscillation mode dominated by DVC loops can shift downward into the EFR in the closed-loop power system. The SCRE is demonstrated in network-mediated IBR plants rather than in a single grid-connected IBR. The SCRE becomes stronger as the scale of IBR plants increases and/or as the minimum eigenvalue of the network admittance matrix decreases.

3) A complex vector-based analysis method is proposed for analyzing DVC-EFOs, from which a stability criterion is derived. The proposed method indicates that system dynamics can be significantly changed by interactions among IBRs as their scale increases, with the network actively participating in this process. Then, EFO mitigation methods are proposed and the critical values of variables are calculated. Consequently, this study identifies an EFO formation mechanism in addition to the traditional EFO mechanisms.

The remainder of this paper is organized as follows. Section II reports an EFO observed in a practical power system with PV plants. Section III establishes the mathematical model of power systems with IBR plants. Section IV clarifies the formation of EFOs caused by the SCRE. In Section V, DVC-EFOs are analyzed, and a stability criterion is derived. In Section VI, the accuracy of the established model and the correctness of the conclusions are verified by EMT simulations. Finally, Section VII summarizes the conclusions.

## II. Report of EFOs in a Practical Power System

### *A. Configuration of the Power System Integrated with PV Plants*

As shown in Fig. 1, the network has a circular topology and is formed by single-circuit 110 kV lines, as indicated by the blue lines. The interconnection points with the city network are marked by green circles, where the voltage level is stepped up to 220 kV. The total installed capacity of the PV plants exceeds 130 MW, and their locations are marked by green blocks. When the EFOs occurred, switch $SW_1$ was open and switch $SW_2$ was closed. Therefore, the PV plants at nodes $S_3$ to $S_5$ exported power to node $T_2$, whereas the PV plants at nodes $S_1$ to $S_2$ exported power to node $T_1$. All PV plants adopt GFL control. The distance from node $S_1$ to $T_1$ is approximately 330 km, while that from node $S_4$ to $T_2$ is approximately 310 km. In particular, the distance between nodes $N_1$ and $T_1$ is 190 km, and that between nodes $N_1$ and $S_2$ is 24 km. The major loads are connected at node $N_1$, while a static var generator is installed at node $N_1$.

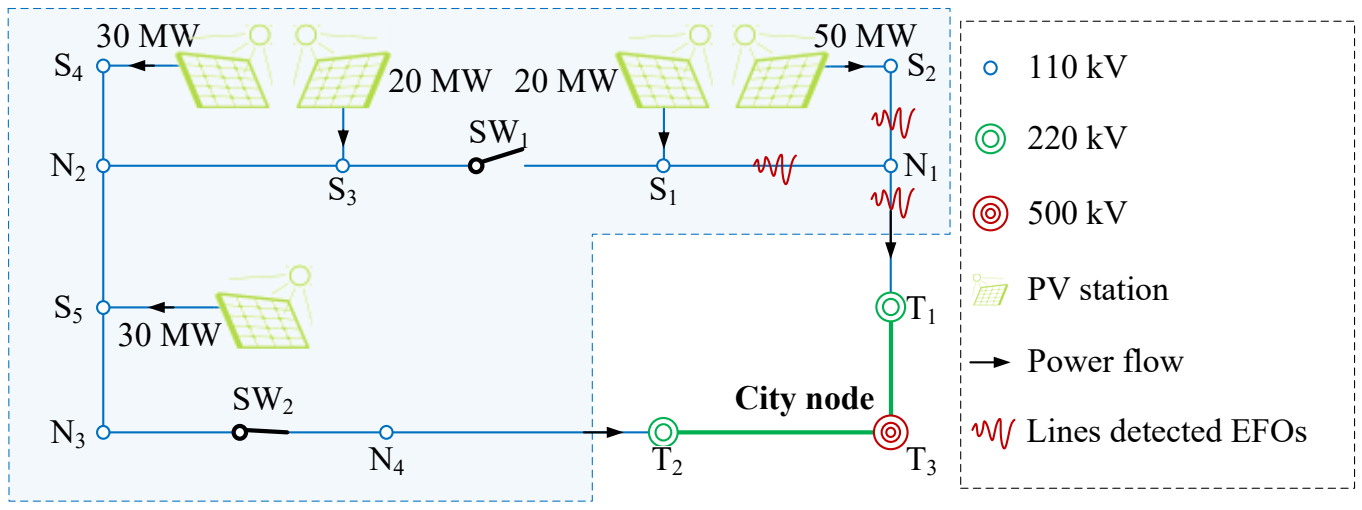

Fig. 1. Configuration of the power system integrated with PV plants.

This power system is in a remote pastoral area far from the city center. There were no local generation sources before PV integration, so residents at $N_1$ were mainly supplied through the 190 km corridor from the urban grid. Because local demand was small, as shown in Table III, voltage problems were not observed in the original operating condition. After 2015, PV deployment in the pastoral area expanded quickly under national support policies to improve electricity access and local livelihoods. PV plants also brought additional income to residents. As a result, the original long distance passive network with light loads gradually evolved into an active network with reverse power flow toward the city center. The reported oscillation occurred after the further expansion of PV capacity at node $S_1$.

This event was associated with multiple factors, including the rapid implementation of national PV-support policies, limited assessment of grid-connection risks and the underdeveloped operation and management practices. Although oscillations under weak grid connections have been widely recognized, especially those related to PLLs, the focus of this paper is to clarify why the oscillations are EFOs rather than conventional subsynchronous oscillations.

## *B. Report of the EFO Caused by PV plants*

The EFO was observed after connecting the PV plants to node $S_1$, that is, as the scale of the PV plants at $S_1$ expanded. The information of the network between switch $SW_1$ and node $T_1$ is listed in Table III.

TABLE III
Information of the PV plants under the normal condition

| Nodes | Sources (MW) | Loads (MW) |
|---|---|---|
| $S_1$ | 20 (Rated power of PV plants) | 0~1 |
| $S_2$ | 50 (Rated power of PV plants) | 0~1 |
| $N_1$ | 0 | 17~26 |

Fig. 2 records the complete process of the observed EFO, including the active and reactive power injected into node $N_1$ from $S_1$ and $S_2$, and the voltage and frequency at node $N_1$. The monitored active power changes gradually rather than through a sudden step or fault. When the active power remains 40 MW, only small fluctuations are observed. As the active power increases from 40 MW toward 50 MW, the EFO becomes evident in the power exchange and is also reflected in both the AC voltage and frequency at node $N_1$. In the time record shown in Fig. 2, the oscillation envelope grows after the active power increase, reaches its largest magnitude near 50 MW. When the active power falls to 30 MW, the oscillation gradually disappears. This record indicates that the oscillation is strongly associated with the PV operation states and the local network condition, rather than with a short-circuit fault or a discrete switching event. The voltage at node $T_3$ remains stable during the event, implying that the oscillation is mainly caused inside the PV sending-end area.

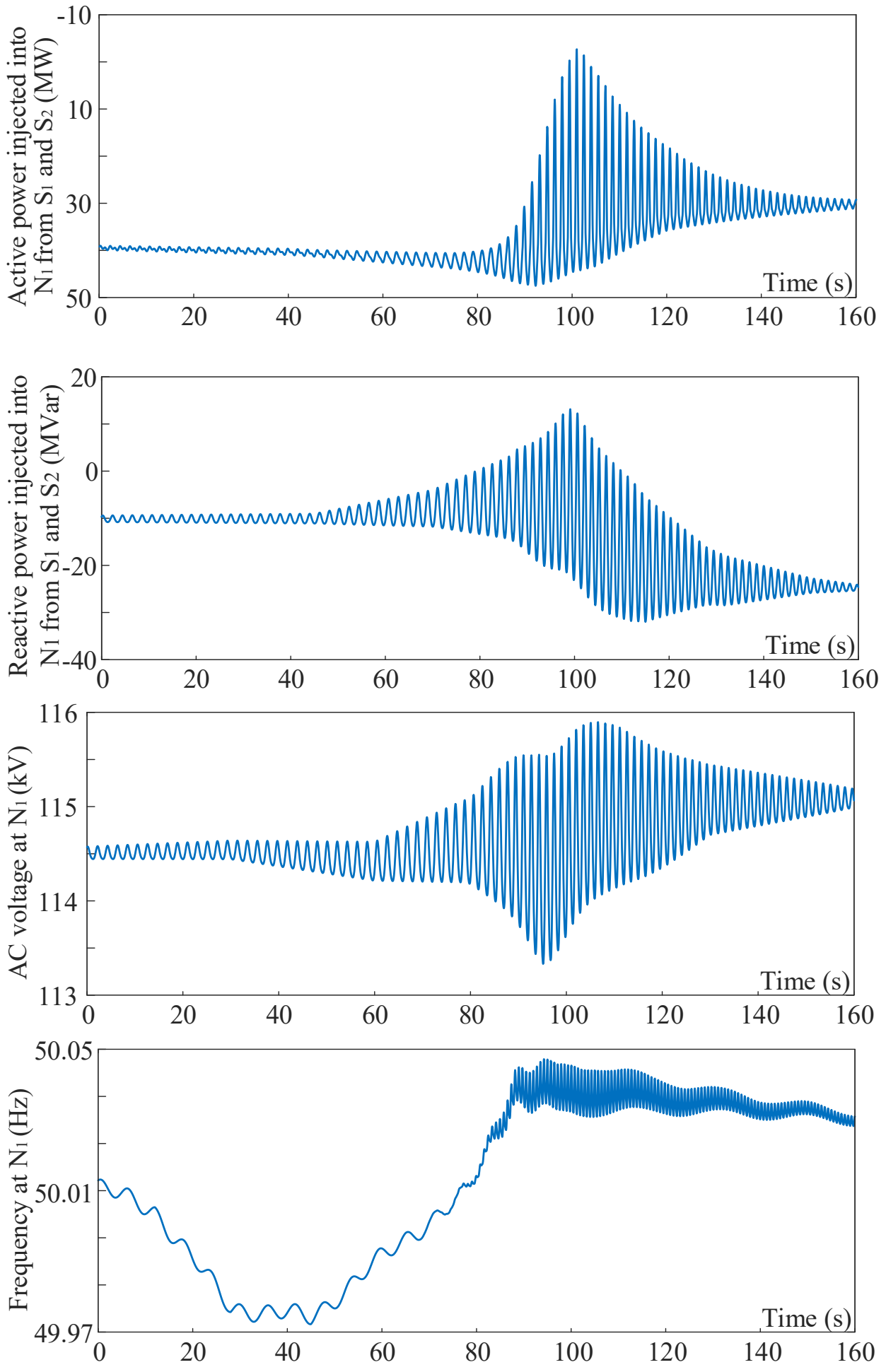

Fig. 2. Monitoring curves during the EFO.

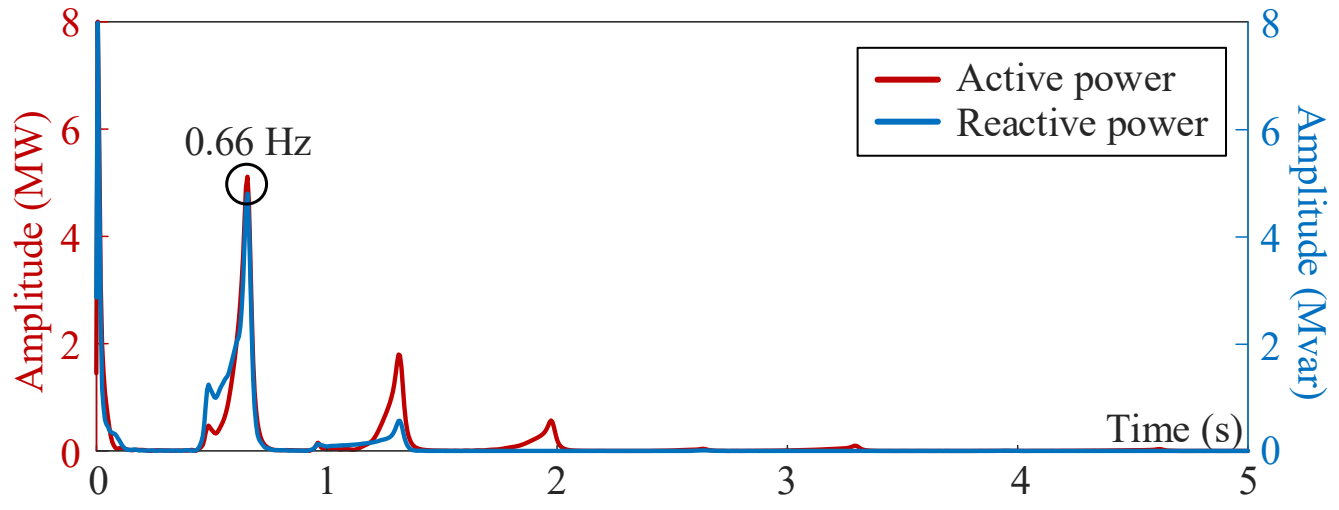

Fig. 3. FFT analysis results of active and reactive power curves.

The FFT results of the active and reactive power curves in Fig. 2 identify an oscillation frequency of 0.66 Hz, as shown in Fig. 3. This frequency falls in the EFR. The record also shows that the EFO is caused by a gradual increase in the output power of PV plants rather than by a sudden fault. Therefore, the event occurs around the system equilibrium point and falls within the scope of small-signal stability analysis.

TABLE IV
Results of practical examinations

| Components | Examinations | Results |
|---|---|---|
| Plant-level control, AGC | Frequent automatic generation control (AGC) instructions may excite frequency oscillations if the commanded power varies repeatedly within a short time. To test this possibility, the AGC instruction was limited during the event. The oscillation nevertheless continued. | Not the cause |
| Plant-level control, AVC | Frequent automatic voltage control (AVC) instructions may produce voltage oscillations through repeated adjustment of reactive power. To test this possibility, the AVC function was withdrawn from operation. The oscillation still persisted. | Not the cause |
| SGs and SG-like IBR controls | EFOs are more commonly associated with SGs or IBRs using SG-like controls. In this event, the oscillation persisted without SGs, and the involved PV inverters used GFL control rather than SG-like control. | Not the cause |

As reviewed in Table I, EFOs may be caused by several mechanisms. Therefore, after the field event was observed, the engineers examined the main candidate sources, as summarized in Table IV. The plant-level control was excluded because the oscillation persisted after AGC and AVC were withdrawn. SGs and SG-like controls were also excluded because there are no SGs or IBRs with SG-like control in the sending-end PV plant, and all IBRs used GFL control. These field observations suggest that the event is related to the PV operation point and local network condition. However, the field record alone is not used as proof that the DVC loop is the only source of the oscillation. Instead, it is used as motivating evidence for the DVC-EFOs studied in this paper. The main question is why an oscillation in the EFR can be caused by GFL-IBRs whose control dynamics are normally much faster.

## III. Mathematical Model of the Power System Integrated with Multiple IBR Plants

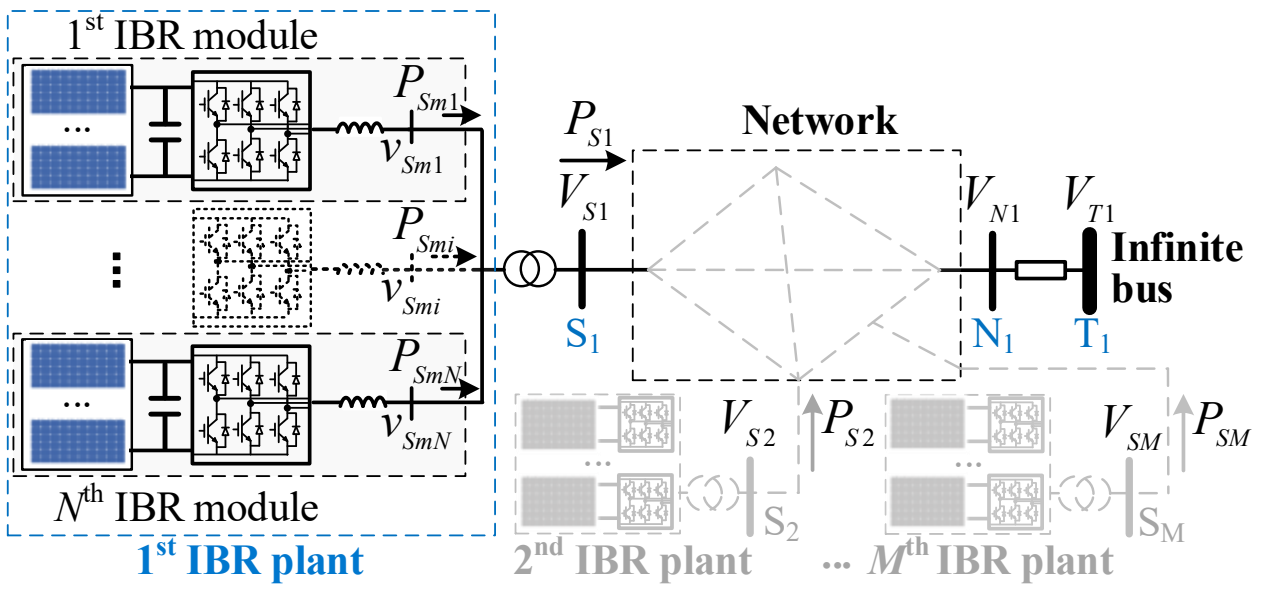

Fig. 4. Configuration of the power system (PV plants are used as an example).

In Fig. 4, $M$ IBR plants are connected to nodes $S_k$, outputting power $P_{Sk}$ to the network, and aggregating at node $N_1$. The external power system is represented by an infinite bus at $T_1$, since this bus did not detect a significant EFO in practice. Each plant contains multiple IBR modules. For example, the plant at node $S_1$ contains $N$ modules, with the output active and reactive power of the $i^{th}$ module denoted as $P_{Smi}$ and $Q_{Smi}$, respectively. The terminal AC voltage and angle are $v_{Smi}$ and $\theta_{Smi}$, respectively. These modules are connected to node $S_1$ through a transformer. $V_{Sk}$ represents the AC voltage of the $k^{th}$ IBR plant, and the angle of $V_{Sk}$ is $\theta_{Sk}$. $V_{Sk} = v_{Smi}$ in per-unit value when the transformer impedance is ignored. To avoid limitations in the analysis, the network topology is generalized and is not restricted to that in Fig. 4. Different plants may contain different numbers of IBR modules, denoted by $N_k$ for the $k^{th}$ IBR plant, to represent their different capacities. For modules in the same plant, the control parameters are assumed to be identical because these modules are usually supplied by the same manufacturer and operated under the same engineering specification. This assumption is reasonable for many practical PV plants, including those in Fig. 1, where modules from the same manufacturer are used for easier management and maintenance. For example, the Templin PV power plant uses 114 SMA Sunny Central 900CP XT inverters [19]. Moreover, module similarity within a PV plant is also a major condition for DVC-EFO occurrence in the practical project. Under this condition, the modules in one plant can behave as a coherent IBR plant and participate in a common network-mediated oscillation mode.

### A. Mathematical Model of the IBR Plant

The DVC is used in the grid-side inverters of IBRs, and the configuration of inner current control (ICC) loops and PLL is illustrated in Fig. 5 [20]. In each IBR plant, $v_{dci}$ and $v_{ci}$ are DC and AC voltages of the $i^{th}$ module, respectively. $l_i$ is the filter inductance on the AC side of the $i^{th}$ module, and $i_{Smi}$ is the AC current flowing through the filter. The capital letter $K$ represents the control parameters, superscript $ref$ indicates the reference value, and subscripts $d$ and $q$ represent the $d$-axis and $q$-axis components of variables, respectively. The introduction of the control loops is not detailed here, as they are well-known to researchers. Since plant-level controls are excluded as the causes in Table IV, the control references of each module are assumed to remain constant during the subsequent derivations.

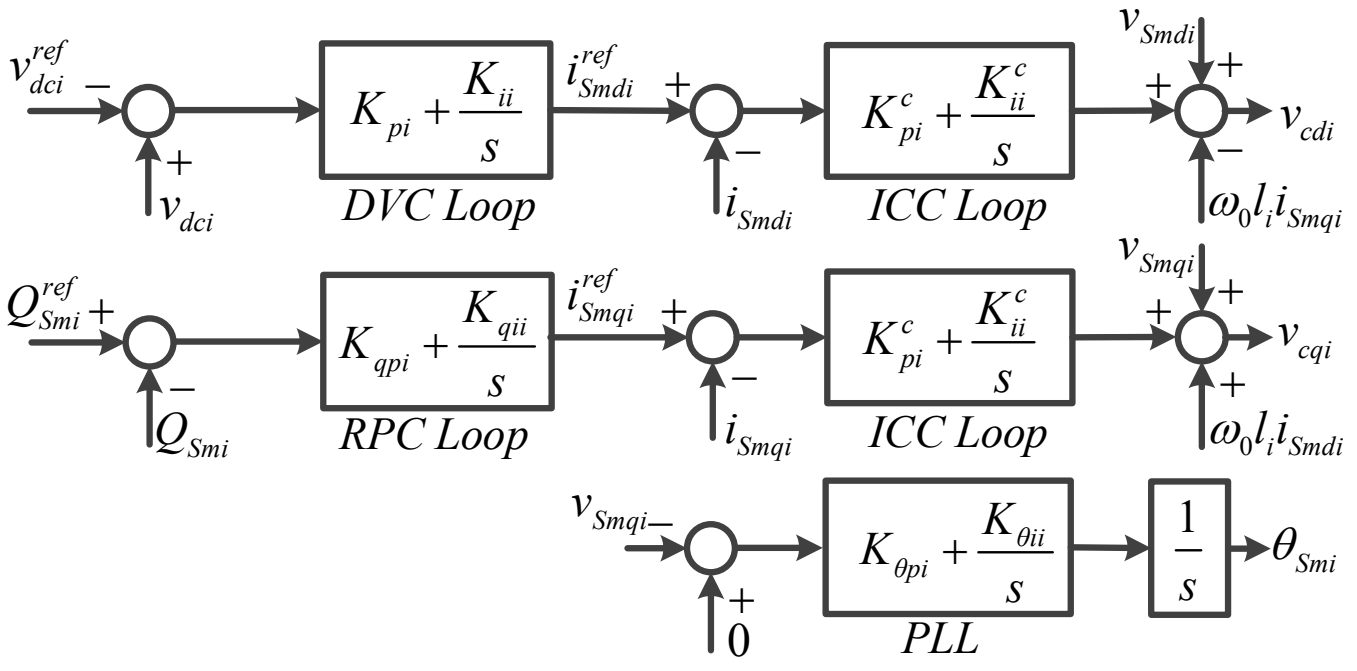

Fig. 5. Control loops of the grid-side inverter of the $i^{th}$ IBR module.

The dynamic model of the grid-side converters has been well documented in [21], [22] and thus will not be detailed in this paper. The governing equations are as follows:

The dynamic equations of the DVC loop:

$$c_i \frac{dv_{dci}}{dt} = \frac{p_{rsi} - (v_{Smdk} i_{Smdi} + v_{Smqi} i_{Smqi})}{v_{dci}}$$
$$i_{Smdi}^{ref} = \left( K_{pi} + \frac{K_{ii}}{s} \right) \left( v_{dci} - v_{dci}^{ref} \right) \tag{1}$$

where $c_i$ is the DC-side capacitance of the grid-side converter and $p_{rsi}$ is the constant active power output from the resource.

The dynamic equations of the ICC loops:

$$i_{Smdi} = G_{ICCi}(s) i_{Smdi}^{ref}, \; i_{Smqi} = G_{ICCi}(s) i_{Smqi}^{ref} \tag{2}$$

where $G_{ICCi}(s) = \dfrac{K_{pi}^c s + K_{ii}^c}{l_i s^2 + K_{pi}^c s + K_{ii}^c}$.

The dynamic equations of the reactive power control (RPC) loop:

$$i_{Smqi}^{ref} = \left( K_{qpi} + \frac{K_{qii}}{s} \right) \left( Q_{Smi}^{ref} - Q_{Smi} \right) \tag{3}$$

By linearizing (1) to (3) at steady state, the transfer function of the $k^{\text{th}}$ IBR plant in *d-q* coordinates can be obtained as

$$\begin{bmatrix} \Delta I_{Sdk} \\ \Delta I_{Sqk} \end{bmatrix} = N_k \begin{bmatrix} g_1(s) & g_2(s) \\ g_3(s) & g_4(s) \end{bmatrix} \begin{bmatrix} \Delta V_{Sdk} \\ \Delta V_{Sqk} \end{bmatrix} \tag{4}$$

where
$$g_1(s) = -\frac{I_{Smdi} G_{ICCi}(s)(K_{pi}s + K_{ii})}{c_i v_{dci} s^2 + v_{Smdi} G_{ICCi}(s)(K_{pi}s + K_{ii})}$$
$$g_2(s) = -\frac{I_{Smqi} G_{ICCi}(s)(K_{pi}s + K_{ii})}{c_i v_{dci} s^2 + v_{Smdi} G_{ICCi}(s)(K_{pi}s + K_{ii})}$$
$$g_3(s) = -\frac{I_{Smqi} G_{ICCi}(s)(K_{qpi}s + K_{qii})}{s + v_{Smdi} G_{ICCi}(s)(K_{qpi}s + K_{qii})}$$
$$g_4(s) = \frac{I_{Smdi} G_{ICCi}(s)(K_{qpi}s + K_{qii})}{s + v_{Smdi} G_{ICCi}(s)(K_{qpi}s + K_{qii})}$$
$$\Delta I_{Sdk} = \sum_{i=1}^{N_k} I_{Smdi}, \Delta I_{Sqk} = \sum_{i=1}^{N_k} I_{Smqi}, \Delta V_{Sqk} = \Delta v_{Smqi}, \Delta V_{Sdk} = \Delta v_{Smdi}$$

The dynamic equations of the PLL:

$$\theta_{Smi} = \left( K_{\theta pi} + \frac{K_{\theta ii}}{s} \right) \left( 0 - v_{Smqi} \right)$$
$$v_{Smqi} = v_{Smi} \sin\left( \theta_{Smi} - \theta_{Pmi} \right) \approx v_{Smi} \left( \theta_{Smi} - \theta_{Pmi} \right) \tag{5}$$

where $\theta_{Pmi}$ is the angle output from the PLL and $\theta_{Smi} = \theta_{Pmi}$ in the steady state.

The transfer function between the *d-q* and *x-y* coordinates is given in (6), which is used to interconnect the transfer functions in (4) with the AC network admittance matrix established in *x-y* coordinates [23].

$$\begin{bmatrix} I_{Sdk} \\ I_{Sqk} \end{bmatrix} = \begin{bmatrix} \cos\theta_{Pk} & \sin\theta_{Pk} \\ -\sin\theta_{Pk} & \cos\theta_{Pk} \end{bmatrix} \begin{bmatrix} I_{Sxk} \\ I_{Syk} \end{bmatrix}$$
$$\begin{bmatrix} V_{Sdk} \\ V_{Sqk} \end{bmatrix} = \begin{bmatrix} \cos\theta_{Pk} & \sin\theta_{Pk} \\ -\sin\theta_{Pk} & \cos\theta_{Pk} \end{bmatrix} \begin{bmatrix} V_{Sxk} \\ V_{Syk} \end{bmatrix} \tag{6}$$

By linearizing (5) and (6) at steady state, the transfer function in *x-y* coordinates can be obtained as

$$\Delta \mathbf{I}_{Sk} = N_k \mathbf{G}_{IBR}(s) \Delta \mathbf{V}_{Sk} \tag{7}$$

where $\mathbf{I}_{Sk} = [I_{Sxk}, I_{Syk}]^T$, $\mathbf{V}_{Sk} = [V_{Sxk}, V_{Syk}]^T$. The *x-y* coordinates are rotating coordinates established on the AC power system, and subscripts *x* and *y* represent the *x*- and *y*-axis variables.

### *B. Mathematical Model of the Power System Connected with M IBR Plants*

The admittance matrix of the network, $\mathbf{Y}_{2M\times 2M}$, can be formed according to the AC transmission lines and loads, as

$$\Delta \mathbf{I}_{2M\times 1} = \mathbf{Y}_{2M\times 2M} \Delta \mathbf{V}_{2M\times 1} \tag{8}$$

where $\mathbf{I}_{2M\times 1} = [\mathbf{I}_{S1}, \ldots, \mathbf{I}_{SM}]^T$, $\mathbf{V}_{2M\times 1} = [\mathbf{V}_{S1}, \ldots, \mathbf{V}_{SM}]^T$.

Therefore, combining (6) with (8), the model of the power system connected with *M* IBR plants can be obtained as

$$Diag\left[ N_k \mathbf{G}_{IBR}(s) \right] = \mathbf{Y}_{2M\times 2M} \tag{9}$$

where *Diag*[ ] denotes a diagonal matrix.

The stability can be determined by the eigenvalues, which are the solutions of (9). However, calculating all oscillation modes directly is time-consuming and cannot provide in-depth insights. The small-signal stability is determined by the oscillation mode with the worst damping, Therefore, the analytical solution for the eigenvalue with the worst damping, known as the dominant oscillation mode, is derived in the following section.

## IV. Formation of DVC-EFOs Caused by the SCRE

### *A. Oscillation Mode Dominated by IBR Plants*

$$\begin{gathered} Diag\left[ N_k \mathbf{G}_{IBR}(s) \right] = \mathbf{Y}_{2M\times 2M} \\ \Downarrow \\ \left( \mathbf{N}\mathbf{E}_{M\times M} \right) \otimes \mathbf{G}_{IBR}(s) = \mathbf{L}_{M\times M} \otimes \mathbf{Y}_0 \\ \Downarrow \\ \mathbf{E}_{M\times M} \otimes \mathbf{G}_{IBR}(s) = \left( \mathbf{N}^{-1} \mathbf{L}_{M\times M} \right) \otimes \mathbf{Y}_0 \\ \Downarrow \\ \left( \mathbf{P}\mathbf{E}_{M\times M} \mathbf{P}^{-1} \right) \otimes \mathbf{G}_{IBR}(s) = \left( \mathbf{P}\mathbf{N}^{-1} \mathbf{L}_{M\times M} \mathbf{P}^{-1} \right) \otimes \mathbf{Y}_0 \\ \Downarrow \\ \mathbf{E}_{M\times M} \otimes \mathbf{G}_{IBR}(s) = Diag\left[ \rho_k \right] \otimes \mathbf{Y}_0 \end{gathered} \tag{10}$$

where $\rho_k$ is the $k^{\text{th}}$ eigenvalue of the matrix $\mathbf{N}^{-1}\mathbf{L}_{M\times M}$, and $\mathbf{P}$ is the left eigenvector matrix of the matrix $\mathbf{N}^{-1}\mathbf{L}_{M\times M}$. $\mathbf{E}_{M\times M}$ is an M-dimensional identity matrix. $\mathbf{Y}_0$ represents the admittance per unit length of a transmission line. $\otimes$ represents the Kronecker product between two matrices.

According to the similarity transformation theory [24], the solutions of (9) are equivalent to those in (10). All oscillation modes can be obtained by solving equations, i.e., $|\mathbf{G}_{IBR}(s)$ - $\rho_k\mathbf{Y}_0|$ = 0. The dominant EFR mode (denoted as $\lambda_E$) is corresponding to the minimum eigenvalue of $\mathbf{N}^{-1}\mathbf{L}_{M\times M}$ [20]. Without loss of generality, denote $\rho_M = \min(\rho_k)$, yields

$$\left|\mathbf{G}_{IBR}(s)-\rho_M\mathbf{Y}_0\right|=0 \xrightarrow{\text{Solution}} s=\lambda_E=-\sigma_E+j\omega_E \tag{11}$$

The minimum eigenvalue of the network admittance matrix represents the weakest network connections among the IBR modules. A smaller minimum eigenvalue means that the same injected current produces a larger voltage deviation across the connected IBR buses. It therefore indicates stronger dynamic interaction through the network. This condition can be caused by long transmission lines, sparse network topology, weak local loads, or an increase in the number of IBR modules sharing the same export corridor. Under these conditions, the network provides less electrical stiffness, and the negative synchronous contribution introduced by SCRE becomes more pronounced.

### *B. Damping and Synchronous Coefficients*

The characteristic equation of the subsystem corresponding to the dominant oscillation mode, that is, (11), can be rewritten as

$$\kappa(s)-\eta(s)=0 \tag{12}$$

where $\kappa(s)$ is a second-order equation that corresponds to the inherent oscillation loop of the IBR, and $\eta(s)$ represents all remaining dynamics.

The expression of $\kappa(s)$ can be written in general as

$$\kappa(s)=s^2+D_s s+T_s \tag{13}$$

where $D_s$ is an inherent damping coefficient and determines whether the oscillation loop $\kappa(s)$ is stable, and $T_s$ is an inherent synchronous coefficient, determining the oscillation frequency.

At the focused frequency point $\omega_E$, the impact of remaining dynamics $\eta(s)$ can be equivalently represented by [25]

$$\bar{\eta}(s)=D_\eta s+T_\eta \tag{14}$$

where $D_\eta=\dfrac{\mathrm{Im}\left[\eta(j\omega_E)\right]}{\omega_E}, T_\eta=\mathrm{Re}\left[\eta(j\omega_E)\right]$.

In (14), $D_\eta$ and $T_\eta$ are the additional damping and synchronous coefficients provided by the remaining dynamics, respectively.

Substituting (13) and (14) into (12), (12) can be equivalently transformed as

$$s^2+\left(D_s+D_\eta\right)s+\left(T_s+T_\eta\right)=0 \tag{15}$$

It can be clearly seen from (15) that the remaining dynamics change the inherent oscillation loop $\kappa(s)$ through an additive term $T_\eta$. In this paper, SCRE is formally defined as a negative equivalent contribution from external subsystems to the oscillation-loop coefficient $T_s$ that causes a downward migration of the oscillation frequency toward the EFR.

### *C. Occurrence of EFOs Caused by the SCRE*

The oscillation frequency of the inherent oscillation loop is in the subsynchronous frequency range, that is

$$\omega_{IN}=\frac{1}{2}\left|\sqrt{D_s^{\,2}-4T_s}\right|\geq\omega_{EMO} \tag{16}$$

where $\omega_{IN}$ is the oscillation frequency of inherent oscillation loop as given by (13), and $\omega_{EFO}$ is the boundary of the EFR (here, the boundary of the EFR is defined as $\omega_{EFO}$ = 2.5 Hz (15.7 rad/s) [26]).

Affected by the SCRE, the oscillation frequency changes to

$$\omega_E=\frac{1}{2}\left|\sqrt{\left(D_s+D_\eta\right)^2-4\left(T_s+T_\eta\right)}\right| \tag{17}$$

The oscillation frequency drops from subsynchronous into the EFR when $0<\omega_E<\omega_{EFO}$, yielding

$$\left(\frac{D_s+D_\eta}{2}\right)^2-T_s<T_\eta<\omega_{EMO}^{\;2}+\left(\frac{D_s+D_\eta}{2}\right)^2-T_s \tag{18}$$

The EFR mode is unstable when

$$D_s+\frac{\mathrm{Im}\left[\eta(j\omega_E)\right]}{\omega_E}<0 \tag{19}$$

## V. Impact of the SCRE on DVC-EFOs

### *A. DVC Oscillation Loop*

For DVC-EFOs, the DVC oscillation loop can be constructed as shown in Fig. 6 [24].

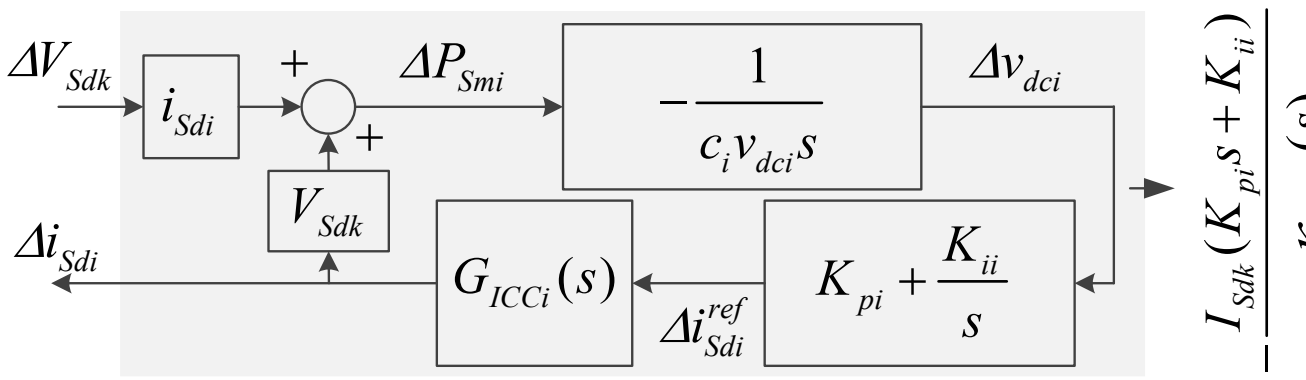

Fig. 6. DVC oscillation loop.

From (5) and Fig. 6, the expression of the second-order DVC oscillation loop can be written as

$$\kappa_{DVC}(s)=s^2+\frac{V_{Sdk}K_{pi}}{v_{dci}c_i}s+\frac{V_{Sdk}K_{ii}}{v_{dci}c_i} \tag{20}$$

From (20), tuning DVC parameters directly affects the open-loop oscillation mode dominated by the DVC loop and changes its inherent damping and frequency. To distinguish this work from studies where EFOs are caused by PLLs or by manual retuning of DVC parameters, two conditions are imposed: i) the open-loop DVC oscillation mode is kept in the subsynchronous frequency range, and ii) PLL parameters are designed with sufficient damping to avoid synchronization instability. Moreover, PLL dynamics are separated from DVC dynamics so that PLL effects on the DVC mode are negligible.

### *B. Complex Vector-based SISO Transfer Function of IBRs*

The mathematical model $\mathbf{G}_{IBR}(s)$ in (7) can always be written in the following form

$$\mathbf{G}_{IBR}(s)=\begin{bmatrix} g_{11}(s) & g_{12}(s) \\ g_{21}(s) & g_{22}(s)\end{bmatrix} \tag{21}$$

$\mathbf{G}_{IBR}(s)$ in (21) can be divided into two parts as

$$\mathbf{G}_{IBR}(s)=\begin{bmatrix} \bar{g}_{11}(s) & \Delta g_{12}(s) \\ \Delta g_{21}(s) & \bar{g}_{11}(s)\end{bmatrix}+\begin{bmatrix} \Delta g_{11}(s) & \bar{g}_{12}(s) \\ \bar{g}_{12}(s) & \Delta g_{22}(s)\end{bmatrix} \tag{22}$$

where

$$\begin{matrix} g_{11}(s)=\bar{g}_{11}(s)+\Delta g_{11}(s) \\ g_{22}(s)=\bar{g}_{11}(s)+\Delta g_{22}(s) \\ \bar{g}_{11}(s)=\left(g_{11}(s)+g_{22}(s)\right)/2 \end{matrix} \quad , \quad \begin{matrix} \bar{g}_{12}(s)=\left(g_{12}(s)+g_{21}(s)\right)/2 \\ g_{12}(s)=\bar{g}_{12}(s)+\Delta g_{12}(s) \\ g_{21}(s)=\bar{g}_{12}(s)+\Delta g_{21}(s) \end{matrix} .$$

In (22), it can be observed that $\Delta g_{11}(s) = -\Delta g_{22}(s)$, $\Delta g_{12}(s) = -\Delta g_{21}(s)$. Therefore (22) equals to

$$\mathbf{I}_{Sk}=\begin{bmatrix} \bar{g}_{11}(s) & -\Delta g_{21}(s) \\ \Delta g_{21}(s) & \bar{g}_{11}(s)\end{bmatrix}\mathbf{V}_{Sk}+\begin{bmatrix} \bar{g}_{12}(s) & -\Delta g_{22}(s) \\ \Delta g_{22}(s) & \bar{g}_{12}(s)\end{bmatrix}\begin{bmatrix} V_{Syk} \\ V_{Sxk}\end{bmatrix} \tag{23}$$

The relation between vectors $V_{Sxk}+jV_{Syk}$ and $V_{Syk}+jV_{Sxk}$ can be expressed as [27]

$$V_{Syk}+jV_{Sxk}=e^{j\theta_k}\left(V_{Sxk}+jV_{Syk}\right) \tag{24}$$

where $\theta_k$ is the angle difference between vectors $V_{Sxk}+jV_{Syk}$ and $V_{Syk}+jV_{Sxk}$.

Based on (22) to (24), the complex vector-based SISO transfer function of the $k^{\text{th}}$ IBR plant can be obtained as

$$\begin{gathered} \vec{I}_{Sk}=N_k p(s)\vec{V}_{Sk} \\ p(s)=\bar{g}_{11}(s)+j\Delta g_{21}(s)+e^{j\theta_k}\left(\bar{g}_{12}(s)+j\Delta g_{22}(s)\right) \end{gathered} \tag{25}$$

where $\vec{I}_{Sk}=I_{Sxk}+jI_{Syk}$, $\vec{V}_{Sk}=V_{Sxk}+jV_{Syk}$, $p_i(s)$ represents the transfer function of the $i^{\text{th}}$ IBR module

*C. Stability Criterion of DVC-EFOs*

Based on (25), (11) changes to

$$p(s)-\vec{\rho}_M=0 \tag{26}$$

where $\vec{\rho}_M$ is a complex value.

Based on (20) and (26), (12) can be specific as

$$\kappa_{DVC}(s)-\frac{1}{\vec{\rho}_M}p_\kappa(s)=0 \tag{27}$$

where, $p(s)$ is divided as $p(s) = p_\kappa(s)/\kappa_{DVC}(s)$.

The condition in (18) can be specific for DVC-EFOs as

$$\begin{gathered} \frac{V_{Sdk}K_{ii}}{2v_{dci}c_i}-\text{Re}\left[\frac{\text{Re}\left[p_\kappa(j\omega_E)\right]+j\,\text{Im}\left[p_\kappa(j\omega_E)\right]}{\text{Re}\left[\vec{\rho}_M\right]+j\,\text{Im}\left[\vec{\rho}_M\right]}\right]\le\omega_{EMO}{}^2 \\ \Downarrow \\ \left|\vec{\rho}_M\right|\left(\frac{V_{Sdk}K_{ii}}{2v_{dci}c_i}-\omega_{EMO}{}^2\right)\le\text{Im}\left[p_\kappa(j\omega_E)\right]\sin\theta_{\rho_M}+\text{Re}\left[p_\kappa(j\omega_E)\right]\cos\theta_{\rho_M} \end{gathered} \tag{28}$$

The condition in (19) can be specific for DVC-EFOs as

$$\begin{gathered} \text{Im}\left[\frac{\text{Re}\left[p_\kappa(j\omega_E)\right]+j\,\text{Im}\left[p_\kappa(j\omega_E)\right]}{\text{Re}\left[\vec{\rho}_M\right]+j\,\text{Im}\left[\vec{\rho}_M\right]}\right]\ge\frac{V_{Sdk}K_{pi}}{2v_{dci}c_i}\omega_E \\ \Downarrow \\ \left|\vec{\rho}_M\right|\frac{V_{Sdk}K_{pi}}{2v_{dci}c_i}\omega_E\le\text{Im}\left[p_\kappa(j\omega_E)\right]\cos\theta_{\rho_M}-\text{Re}\left[p_\kappa(j\omega_E)\right]\sin\theta_{\rho_M} \end{gathered} \tag{29}$$

By squaring equations in (28) and (29), and then combining them, the DVC-EFOs occurs if

$$\left|\vec{\rho}_M\right|\le\left|p_\kappa(j\omega_E)\right|/C_0 \tag{30}$$

where, $C_0=\left(\frac{V_{Sdk}K_{ii}}{2v_{dci}c_i}-\omega_{EMO}{}^2\right)^2+\left(\frac{V_{Sdk}K_{pi}}{2v_{dci}c_i}\omega_E\right)^2$.

Therefore, a smaller eigenvalue of the network admittance matrix is a significant index for DVC-EFOs, which is unrelated to the control parameters but is related to the scale of IBRs and configuration of the network.

*D. Impact Factors of the SCRE*

In this section, the scale of IBR plants, network topologies, and network loads are analyzed. The detailed impacts of control parameters are not covered in this paper to differ from other EFOs caused by tuning control parameters.

If the IBR plants expand, the number of IBR modules in the $k^{\text{th}}$ IBR plant increases from $N_k$ to $\alpha N_k$, where $\alpha > 1$. Thus, (10) changes to

$$\begin{gathered} Diag\left[\alpha N_k\mathbf{G}_{IBR}(s)\right]=\mathbf{Y}_{2M\times 2M} \\ \Downarrow \\ \mathbf{E}_{M\times M}\otimes\mathbf{G}_{IBR}(s)=\alpha^{-1}Diag\left[\rho_k\right]\otimes\mathbf{Y}_0 \end{gathered} \tag{31}$$

Consequently, (26) changes to

$$p(s)-\alpha^{-1}\vec{\rho}_M=0 \tag{32}$$

The stability criterion in (30) modifies to

$$\left|\vec{\rho}_M\right|\le\alpha^{-1}\left|p_\kappa(j\omega_E)\right|/C_0 \tag{33}$$

It can be concluded that as the scale of IBR plants increases, the stable region will be reduced, thus increasing the possibility of satisfying the conditions listed in (33). Therefore, expanding the scale of IBR plants poses a risk to stability. The maximum scale of IBRs should be limited.

If an additional line is added between the $i^{\text{th}}$ and $j^{\text{th}}$ nodes, the admittance matrix changes from (8) to

$$\vec{\mathbf{I}}_{M\times 1}=\left(\mathbf{Y}_{M\times M}+\mathbf{Y}_{Line}\right)\vec{\mathbf{V}}_{M\times 1} \tag{34}$$

where $y_{ij}$ is the admittance of the added line, and

$$\mathbf{Y}_{Line}=\begin{bmatrix}\ddots & \vdots & & \vdots & \\ \cdots & y_{ij} & \cdots & -y_{ij} & \cdots \\ & \vdots & \ddots & \vdots & \\ \cdots & -y_{ji} & \cdots & y_{ji} & \cdots \\ & \vdots & & \vdots & \ddots\end{bmatrix}.$$

We denote the minimum eigenvalue of $(\mathbf{Y}_{M\times M}+\mathbf{Y}_{Line})$ as $\vec{\rho}_{ij}$, it can be verified that

$$\begin{aligned}\left|\vec{\rho}_{ij}\right| &\approx \left|\mathbf{v}\mathbf{Y}_M\mathbf{v}^T+\mathbf{v}\mathbf{Y}_{Line}\mathbf{v}^T\right| \\ &=\left|\vec{\rho}_M+{v_i}^2y_{ij}-v_jv_iy_{ij}-v_iv_jy_{ij}+{v_j}^2y_{ij}\right| \\ &=\left|\vec{\rho}_M+y_{ij}\left(v_i-v_j\right)^2\right|\geq\left|\vec{\rho}_M\right|\end{aligned} \tag{35}$$

where $\mathbf{v}$ is the characteristic eigenvalue vector corresponding to $\vec{\rho}_M$, and $v_i$ and $v_j$ are the $i^{th}$ and $j^{th}$ elements of $\mathbf{v}$.

It can be seen that adding a line to network can increase the minimum eigenvalue of admittance matrix, thereby reducing the possibility of satisfying the conditions listed in (30). Therefore, adding lines to the network can always improve stability.

Increasing the active power load is equivalent to reducing the resistance connected to the network. Similar to (33), if the active power loads at node $k$ increase, the admittance matrix changes from (34) to

$$\vec{\mathbf{I}}_{M\times 1}=\left(\mathbf{Y}_{M\times M}+\mathbf{Y}_{Load}\right)\vec{\mathbf{V}}_{M\times 1} \tag{36}$$

where, the $k^{th}$ row and the $k^{th}$ column element of $\mathbf{Y}_{Load}$ is $y_{kk}$. $y_{kk}$ is the changed admittance at node $k$, and is a real number with a positive value when the active power increases.

A similar derivation to (34) can be conducted, and it can be concluded that increasing the network loads can increase the real part value of the minimum eigenvalue of admittance matrix, thereby reducing the possibility of satisfying conditions listed in (30). Therefore, promoting customer use of active power can improve stability.

Therefore, based on these analyses, the SCRE can be reduced from the perspectives of reducing sources, enhancing networks, and increasing the local loads of customers.

## VI. Case Study

### A. Model Verifications

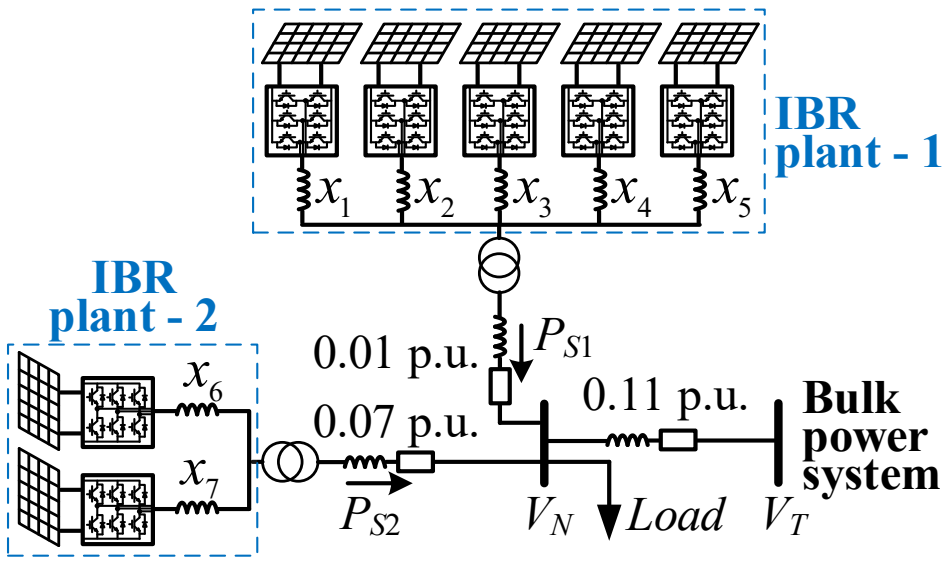

Fig. 7. A power system connected with two IBR plants.

As shown in Fig. 7, the test system contains two IBR plants, with five IBR modules in one plant and two in the other. PV plants are used as representative IBRs. All modules are connected in parallel to the common bus, adopt the converter-level control structure in Fig. 5, and do not participate in AVC or AGC adjustments. The system is implemented in the EMT simulation platform Simulink. The parameters are listed in Table V in Appendix A.

The mathematical equations of the power system in Fig. 7 can be obtained using a MATLAB program, and the numerical expression of (26) is derived when the output power of each PV module is 0.72 p.u., as

$$\begin{aligned}p(s)&=\frac{(0.8379s+0.2445)(-0.4634-j0.4987)}{5.0s^2+0.8486s+0.2476} \\ &\quad+0.5298-j0.4924 \\ \vec{\rho}_M&=0.0670-j0.9902\end{aligned} \tag{37}$$

In (37), the poles of $p(s)$ are calculated as $-0.08486\pm j0.2057$ (p.u.), i.e., $-26.66\pm j64.63$ (rad/s). Therefore, the inherent frequency of the IBR module is 10.29 Hz.

Considering the dynamic interactions among IBRs through the network, the EFR mode can be determined by substituting (37) into (26), yielding $\lambda_E=-0.00093\pm j0.02332$ (p.u.), i.e., $-0.29\pm j7.33$ (rad/s). This indicates that EFO occurs with a frequency of 1.17 Hz when each IBR module outputs a power of 0.72 p.u.

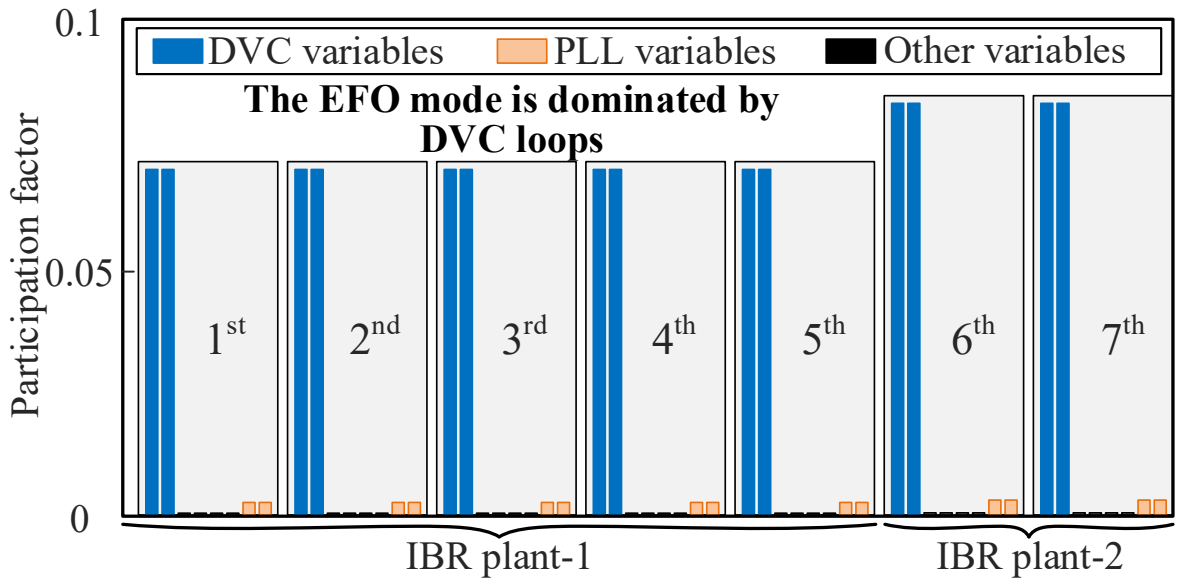

Fig. 8. Participation factor of the EFR mode.

The participation factors of the EFR mode are shown in Fig. 8. For each IBR module, seven state variables are considered in the following order: the DC-link voltage $v_{dci}$, the DVC integral state, the $d$- and $q$-axis currents $i_{Smdi}$ and $i_{Smqi}$, the current-control integral states , the PLL angular speed, and the PLL integral state. The first two variables belong to the DVC loop, the last two variables belong to the PLL, and the remaining variables correspond to the ICC loops. The summed participation of the DVC-related states reaches 97%, whereas the PLL and ICC states account for only 2% and 1%, respectively. This distribution confirms that the EFR mode is dominated by the DVC loops. The PLL and ICC loops are weakly involved under the selected parameters and do not constitute the primary source of the observed EFO.

An EMT simulation is then conducted in Simulink. Because all IBR modules operate under similar external conditions, the output power of the first plant is approximately 2.5 times that of the second plant. Fig. 9(a) shows the output active power of each IBR module, Fig. 9(b) shows the system frequency, and Fig. 9(c) shows the FFT results. Two scenarios are compared. The first scenario contains two IBR plants with seven PV modules, each operating at 0.72 p.u. This case includes network-mediated interactions among modules and is shown by the red curves. The second scenario contains only one IBR module operating at 0.72 p.u. This case represents the inherent module dynamics without network-mediated interactions and is shown by the blue curves.

In both scenarios, the system is stable before the disturbance. A 1.4% active power disturbance is applied at 1 s and cleared at 1.02 s. In the seven-module case, the red curves show a slow response with weak damping, and the FFT result identifies an oscillation frequency of 1.34 Hz. This frequency is close to the calculated EFO frequency obtained from (37). In contrast, the single-module case shows a faster and more strongly damped response. Its oscillation frequency is 10.25 Hz, which agrees with the inherent DVC frequency calculated from (37). The comparison validates the mathematical model and confirms that the EFO appears only after the IBR modules are strongly coupled through the network.

The main stability concern is not only whether the response becomes unstable, but also why a DVC-dominated oscillation mode can move into the EFR. Increasing the IBR scale increases the total output active power, as in the field event, but the critical point here is the frequency migration. The single-module case retains the inherent DVC oscillation frequency around 10.25 Hz, outside the EFR, whereas the seven-module case shifts the closed-loop mode to 1.34 Hz in the EFR. Because the PLL is well damped in the study case, the EFO is attributed to the DVC loop and network-mediated interactions rather than to PLL synchronization instability.

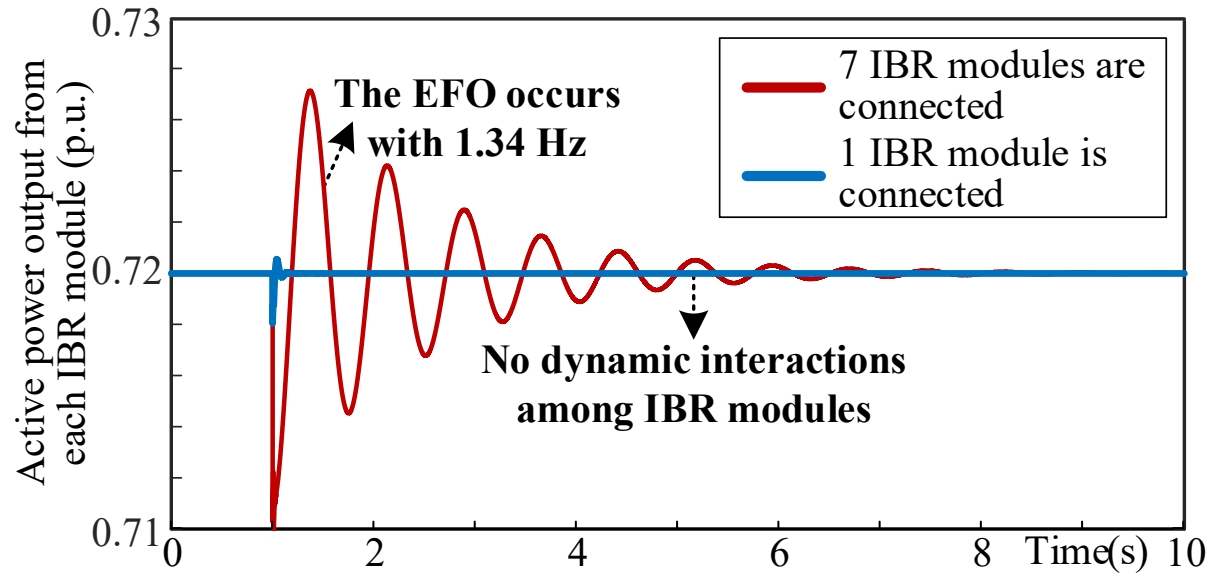

(a) Output active power of each IBR module.

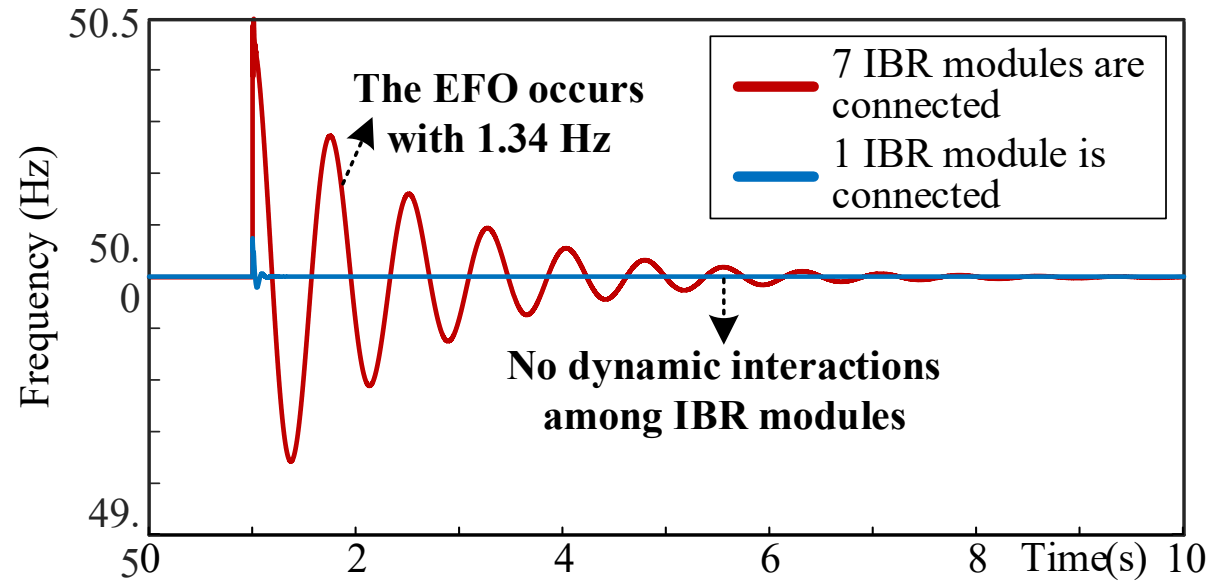

(b) Frequency of the power system.

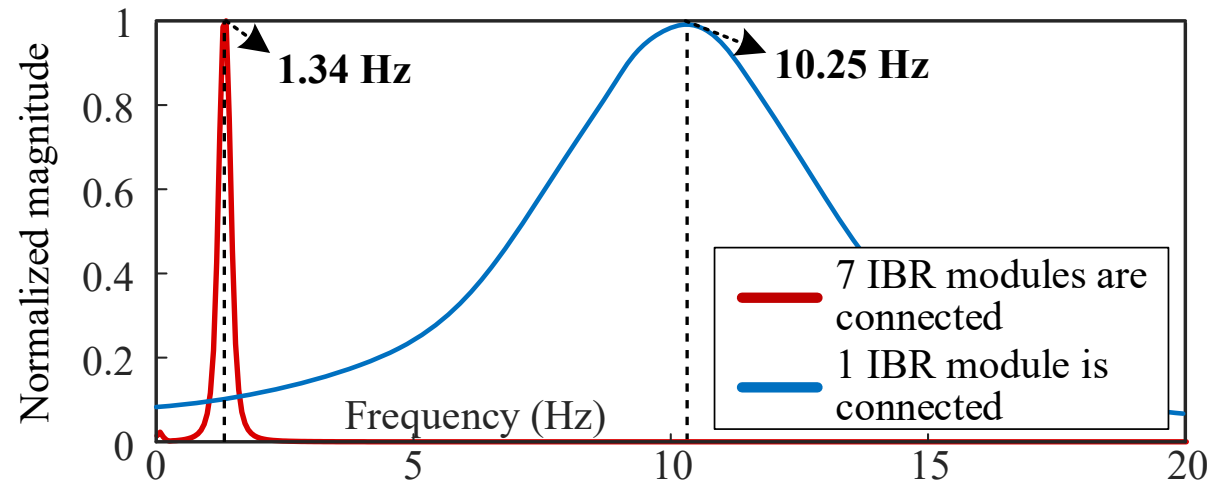

(c) FFT analysis results.

Fig. 9. Simulation results when each IBR module outputs power of 0.72 p.u.

*B. Impact of Different Inherent Oscillation Frequency on DVC-EFOs*

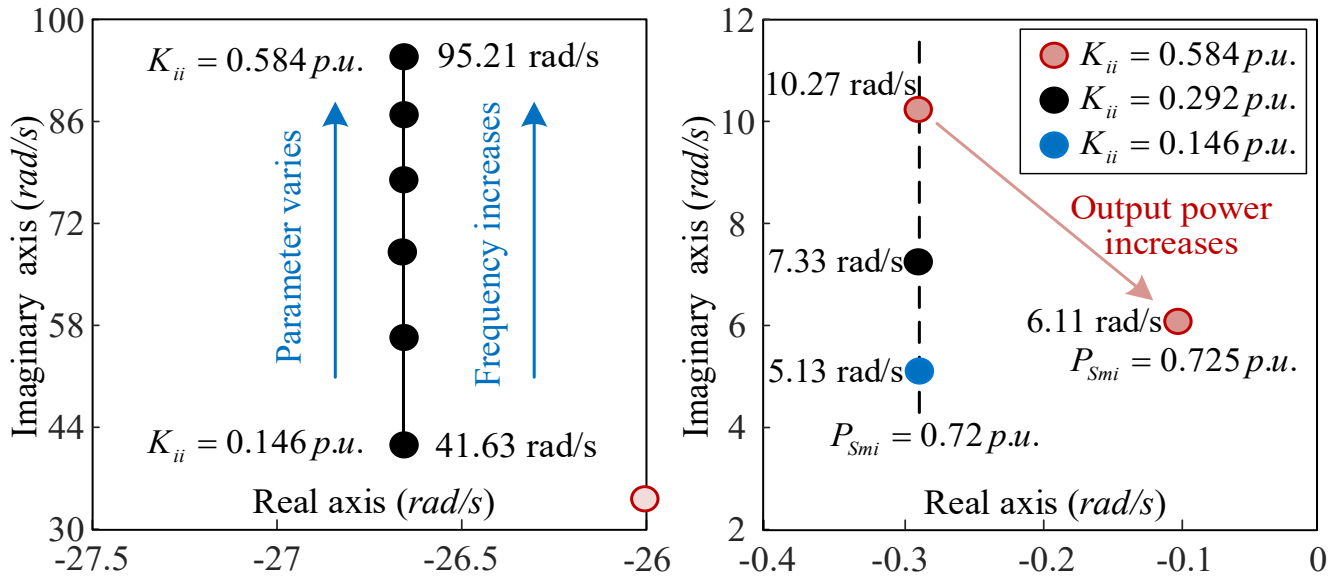

(a) Open-loop DVC oscillation mode (b) Closed-loop DVC oscillation mode

Fig. 10. Trajectory of open-loop DVC oscillation mode and $\lambda_E$ under different DVC control parameters.

The open-loop DVC oscillation mode is used to represent the inherent DVC dynamics of an IBR module. Its frequency can be evaluated from the imaginary part of the open-loop mode. Fig. 10 compares the open-loop and closed-loop DVC oscillation modes as the DVC integral gain $K_{ii}$ varies from 0.146 p.u. to 0.584 p.u. The open-loop DVC frequency increases from 41.63 rad/s to 95.21 rad/s, namely from 6.6 Hz to 15.2 Hz. These frequencies remain above the EFR boundary. However, after the modules are coupled through the network, the corresponding closed-loop frequencies decrease to 10.27, 7.33, and 5.13 rad/s when each PV module outputs 0.72 p.u. The closed-loop modes therefore enter the EFR even though the inherent DVC modes do not. A higher open-loop frequency still produces a higher closed-loop frequency under otherwise identical conditions, but the closed-loop frequency can continue to decrease when the exported active power increases. For example, increasing the output power from 0.72 p.u. to 0.725 p.u. reduces the closed-loop frequency from 10.27 rad/s to 6.11 rad/s. These results support the proposed mechanism: DVC-EFOs are caused by network-mediated interactions among IBRs, which create the SCRE and pull the closed-loop DVC mode into the EFR. This differs from traditional EFO mechanisms in which the open-loop and closed-loop modes are already located in or close to the EFR.

*C. Impact of RPC Loop on DVC-EFOs*

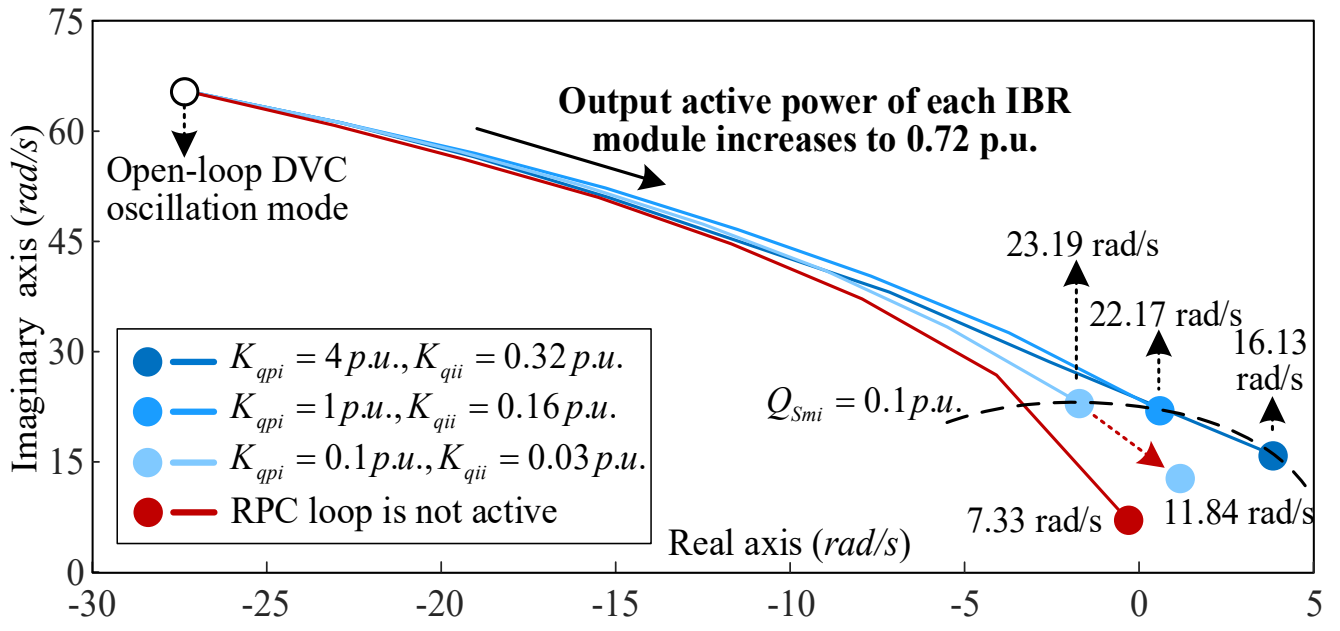

Fig. 11. Trajectory of $\lambda_E$ under different reactive power outputs.

In practical operation, PV plants often operate near full active power output. When the RPC loop is enabled, an IBR module can also provide reactive power support, which changes the operating point and the network-mediated dynamics. In Fig.

11(a), the RPC reference increases from 0 p.u. to 0.1 p.u. The oscillation frequency recovers from the condition without reactive power, indicating that RPC can weaken the SCRE. The degree of improvement depends on the RPC parameters. The frequency can be increased to at least 16.13 rad/s in the examined cases, but damping is not always improved. The damping remains negative for RPC gains of [0.1, 0.03] and [1, 0.16], whereas it becomes positive for [4, 0.32]. Therefore, RPC affects the SCRE but does not guarantee stability. Moreover, RPC cannot eliminate the DVC-EFO mechanism fundamentally. When each IBR module outputs 0.72 p.u. active power, increasing reactive power from 0 p.u. to 0.1 p.u. raises the oscillation frequency from 7.33 rad/s to 23.19 rad/s. However, when the active power output further increases to 0.79 p.u., the frequency decreases from 23.19 rad/s to 11.84 rad/s.

*D. Critical Number of IBR modules*

To ensure that all generated power can be transmitted out of the sending-end region, an IBR model with rated power and no residential load is used in the analysis. This setup enables the determination of the critical number of IBR modules for stable operation. The mathematical model of an IBR module operating at rated power is given by

$$p(s)=\frac{(1.175s+0.3428)(-0.4285-j0.4949)}{5.0s^2+0.8511s+0.2483}+0.7889-j0.6831 \tag{38}$$

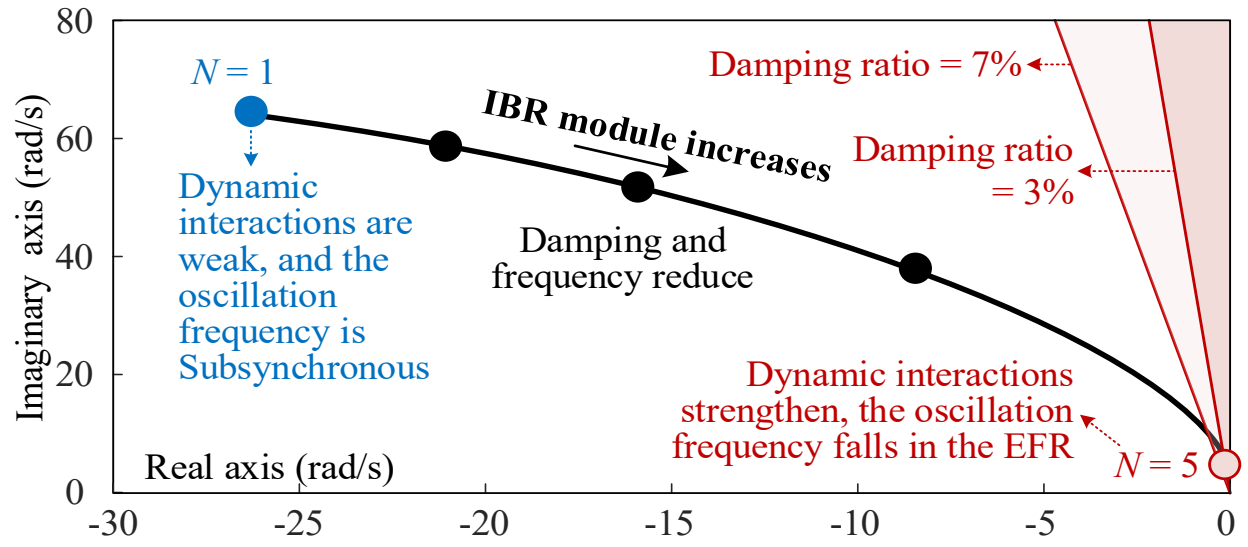

Fig. 12. Trajectory of $\lambda_E$ when the IBR module number increases.

As the total number of IBR modules increases from $N = 1$ to $N = 5$, the trajectory of the closed-loop EFR mode is shown in Fig. 12. When $N = 1$, the oscillation mode is -26.55 + $j$64.52 rad/s, as indicated by the blue circle. This mode corresponds to a subsynchronous inherent frequency of 10.27 Hz and has strong damping. As the number of IBR modules increases, the dynamic interactions among IBRs strengthen, and both oscillation damping and frequency decrease along the black trajectory. When $N = 5$, the EFR mode shifts to -0.12 + $j$4.66 rad/s, as indicated by the red circle. The oscillation frequency then drops to 0.74 Hz, placing the mode in the EFR with weak damping.

During this transition, the inherent dynamics of each module remain unchanged, that is, the poles of (38) remain fixed. However, $\vec{\rho}_M$ changes as the number of IBR modules increases, as illustrated in Fig. 13. As more modules are integrated, the magnitude of $\vec{\rho}_M$ decreases. Together with Fig. 12, this confirms that increasing the number of IBR modules degrades system stability and shifts the oscillation mode toward lower frequency and lower damping by reducing $\vec{\rho}_M$.

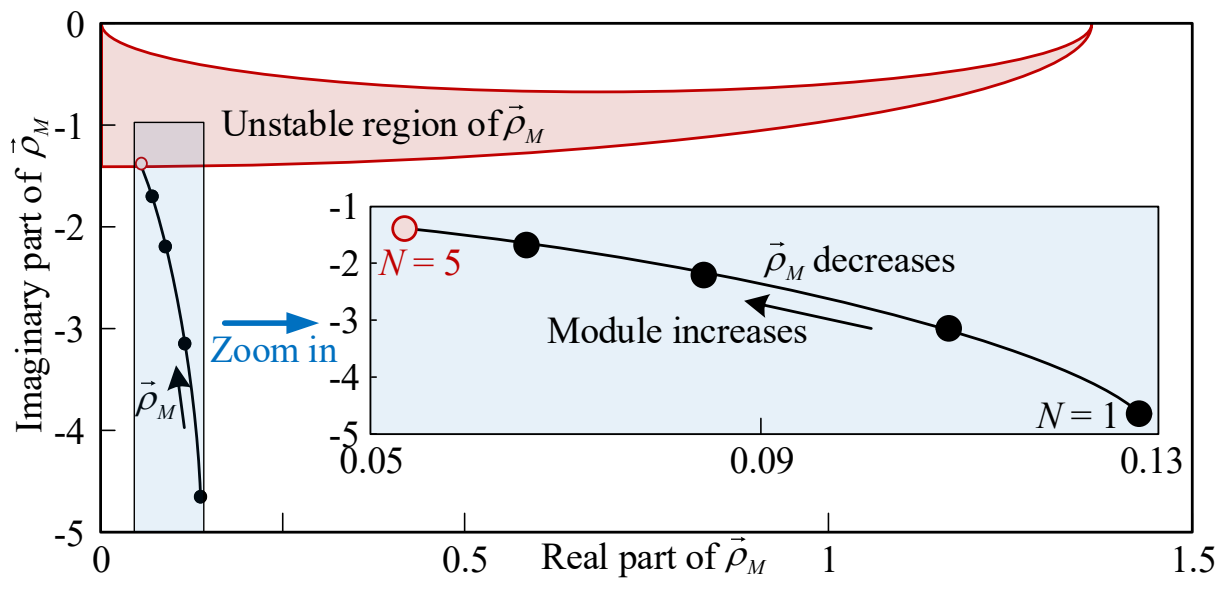

Fig. 13. Trajectory and unstable region of $\vec{\rho}_M$.

To determine the critical number of IBR modules, the unstable region of $\vec{\rho}_M$ is calculated using the conditions in (30) and the characteristic equation $p(s)$ in (38). The results are highlighted in red in Fig. 13. When the number of IBR modules ranges from 1 to 4, $|\vec{\rho}_M|$ remains above the stability threshold, ensuring stable operation. However, when the number of IBR modules reaches 5, $|\vec{\rho}_M|$ falls into the unstable region and drives the system into the EFR with insufficient damping. This result confirms that lower network admittance reduces values in both the real and imaginary components of $\vec{\rho}_M$, and the allowable IBR module number. The critical stability threshold is identified as $|\vec{\rho}_M| = 1.37$. Specifically, if $|\vec{\rho}_M| > 1.37$, the system is stable. However, if $|\vec{\rho}_M| < 1.37$, the system becomes susceptible to instability.

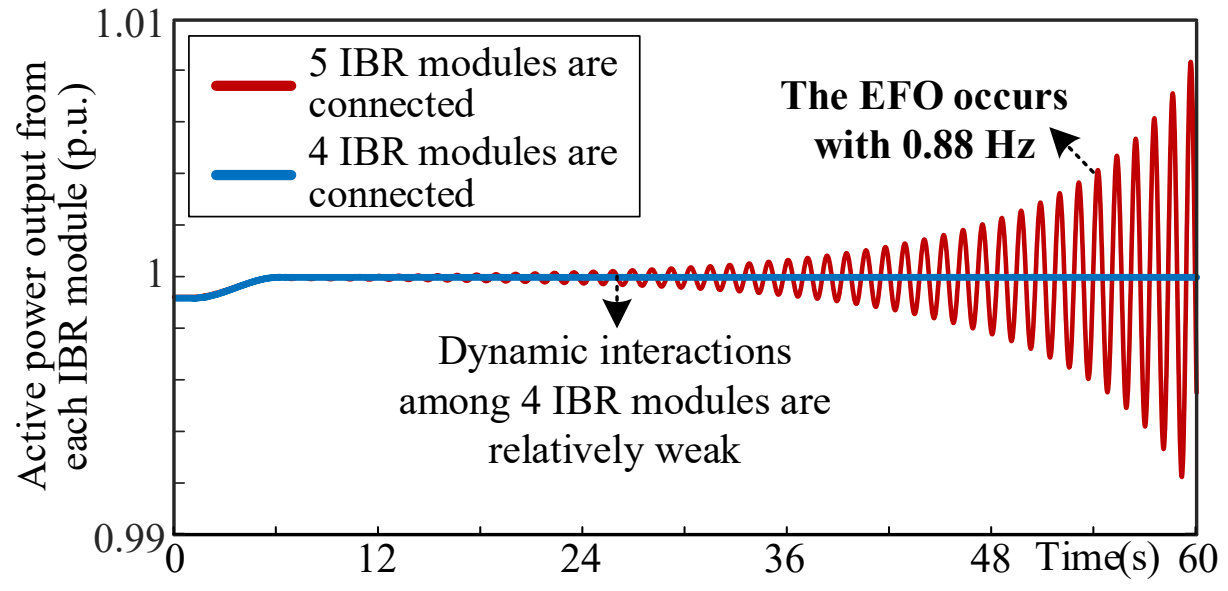

(a) Output active power of each IBR module.

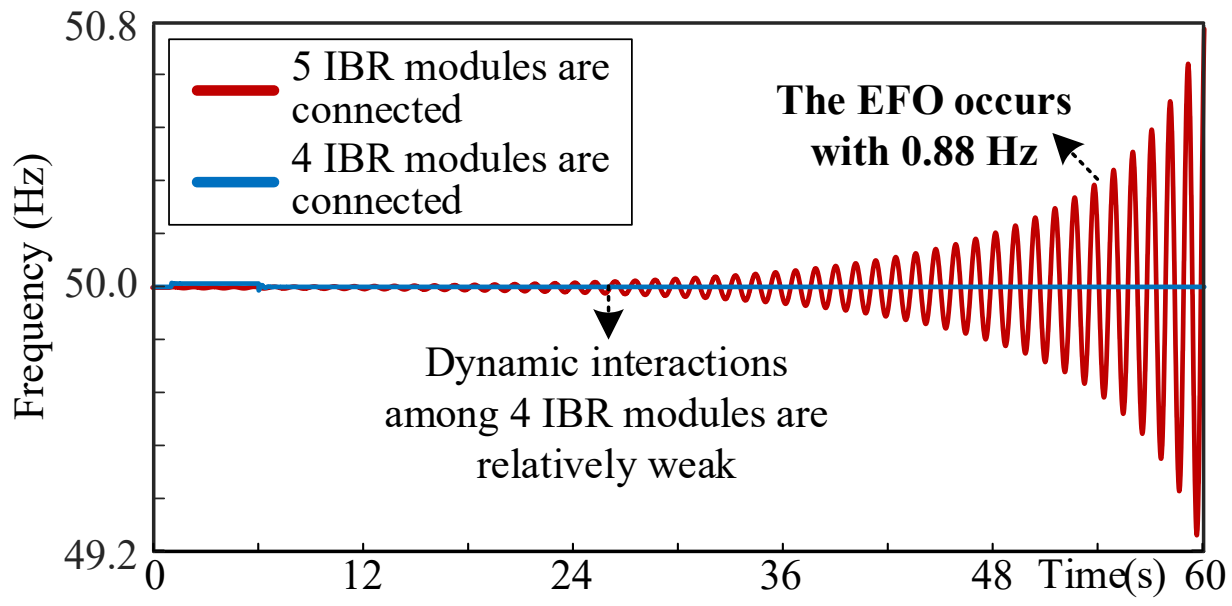

(b) Frequency of the power system.

Fig. 14. Simulation results when IBR module number increases.

Fig. 14 verifies the analysis by comparing cases with four and five IBR modules. In both cases, the system is initially stable. At 1 s, the active power reference of each module is stepped to 1 p.u., and the active power slowly increases. With four modules, the active power and frequency responses remain bounded and well damped, indicating that the system can transmit rated power at this scale. With five modules, the red curves show a growing EFO at approximately 0.88 Hz, and no short-circuit fault or switching event is applied during the oscillation. This behavior is consistent with the practical EFO record and confirms that the original network cannot support five fully loaded IBR modules without additional mitigation. Therefore, the maximum allowable IBR scale must be assessed together with the network admittance condition, rather than by converter rating alone.

## *E. Critical Line Length of Network*

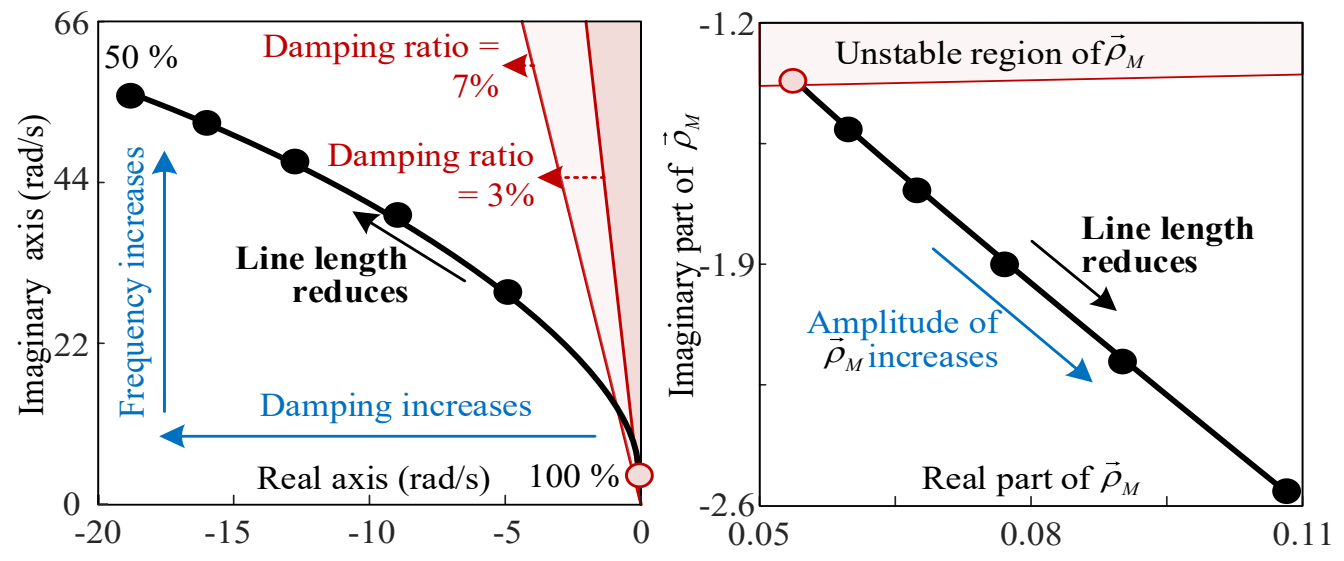

Fig. 15. Trajectories of $\lambda_E$ and $\vec{\rho}_M$ when the line length decreases.

Maintaining the IBR module number at $N = 5$, the line length is reduced by 50% from the original value. The trajectories of $\lambda_E$ and $\vec{\rho}_M$ are shown in Fig. 15. As shown in Fig. 15, reducing the line length improves stability. This improvement is attributed to the increased admittance, which raises $\vec{\rho}_M$ when the line length is reduced. This result shows that increasing transmission admittance, or equivalently reducing line impedance, improves the ability of the network to accommodate more IBR modules. Weak network connections therefore pose a higher risk of DVC-EFOs. Each circle in Fig. 15 corresponds to a 10% reduction in line length from the original value. Using 7% damping ratio requirement as the stability constraint, the full-power delivery condition for all IBR modules can be satisfied when the effective line length is reduced by 3%. At this operating point, the dominant mode is -0.93 + $j$13.2 rad/s.

However, reducing the line length may be impossible because of geographic constraints. An equivalent mitigation is to install an additional line in parallel with part of the existing lines. For example, paralleling an identical line over the current one can halve the equivalent impedance, which is equivalent to reducing the line length by 50%. If an identical parallel line is added over $T$ % of the original corridor, the impedance reduction can be evaluated as $\Delta z_L$ = -0.5$z_L T$ %, where $z_L$ is the impedance of original lines.

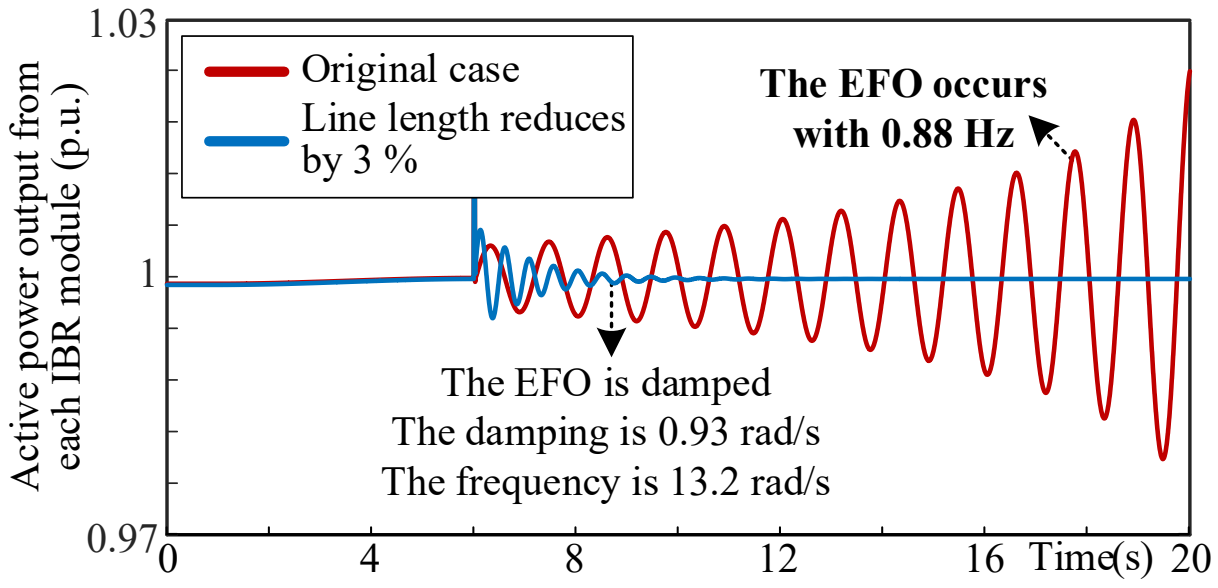

Fig. 16. Output active power of each IBR module when line length decreases.

To validate this result, EMT simulations are performed as shown in Fig. 16. At 1 s, the active power reference of each module is stepped to 1 p.u., and a 2% active power disturbance is applied at 6 s and cleared at 6.02 s. When the line length is effectively reduced by 3%, the active power responses of all IBR modules remain well damped after the disturbance. The simulation confirms the results from Fig. 15: a proper increase in line admittance can move the dominant oscillation mode outside the weak damping region and maintain stable full-power delivery.

## *F. Critical Local Loads in Power System*

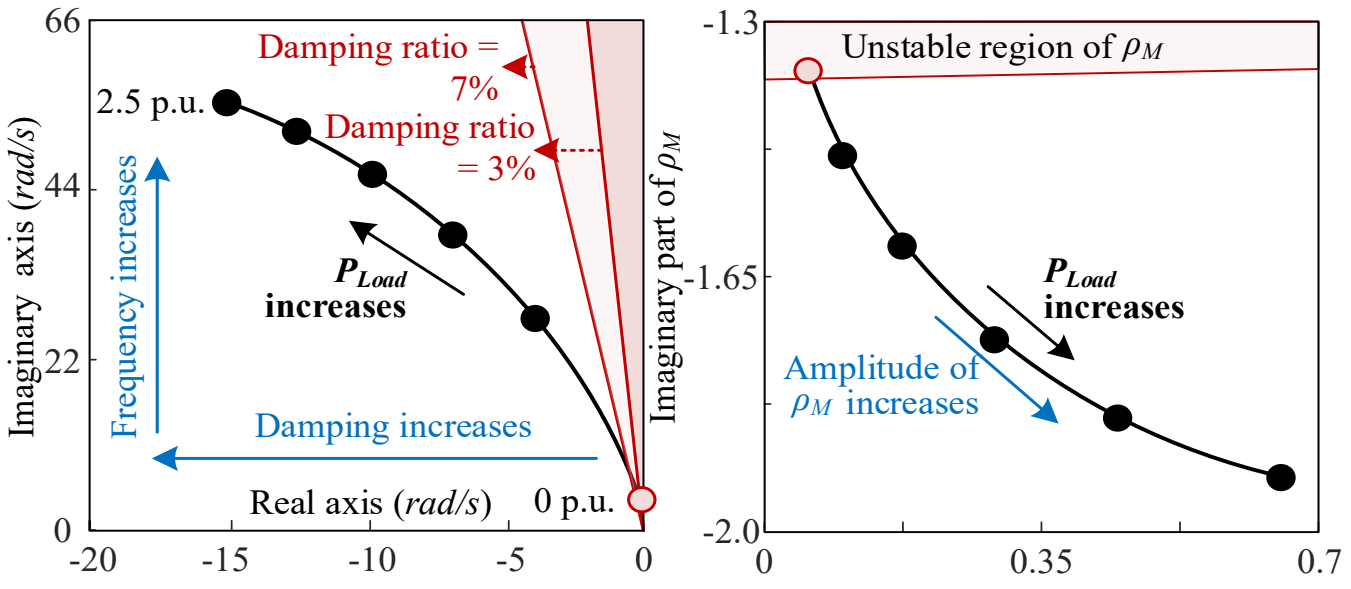

Fig. 17. Trajectory of $\lambda_E$ and $\vec{\rho}_M$ when load power increases.

Maintaining $N$ = 5 and the line parameters unchanged, the local load power in Fig. 7, denoted as $P_{Load}$, is increased from 0 p.u. to 2.5 p.u. The trajectories of $\lambda_E$ and $\vec{\rho}_M$ are shown in Fig. 17. Fig. 17 shows that increasing local load improves stability because the magnitude of $\vec{\rho}_M$ increases as the active load grows. Thus, increasing local active power consumption strengthens the network admittance seen by the IBR plants and allows more IBR modules to operate stably. Each circle in Fig. 17 corresponds to a 0.5 p.u. increase in load power. With the 7% damping-ratio requirement imposed as the stability constraint, full-power delivery from the IBR modules requires at least 0.06 p.u. of local active load. In this case, the dominant oscillation mode is -0.54 + $j$7.7 rad/s. Because residential load cannot always be guaranteed to exceed 0.06 p.u., an energy-storage system or controllable load can be used to absorb part of the generated power when local demand is insufficient. The combined residential load and storage charging power should therefore remain no less than 0.06 p.u.

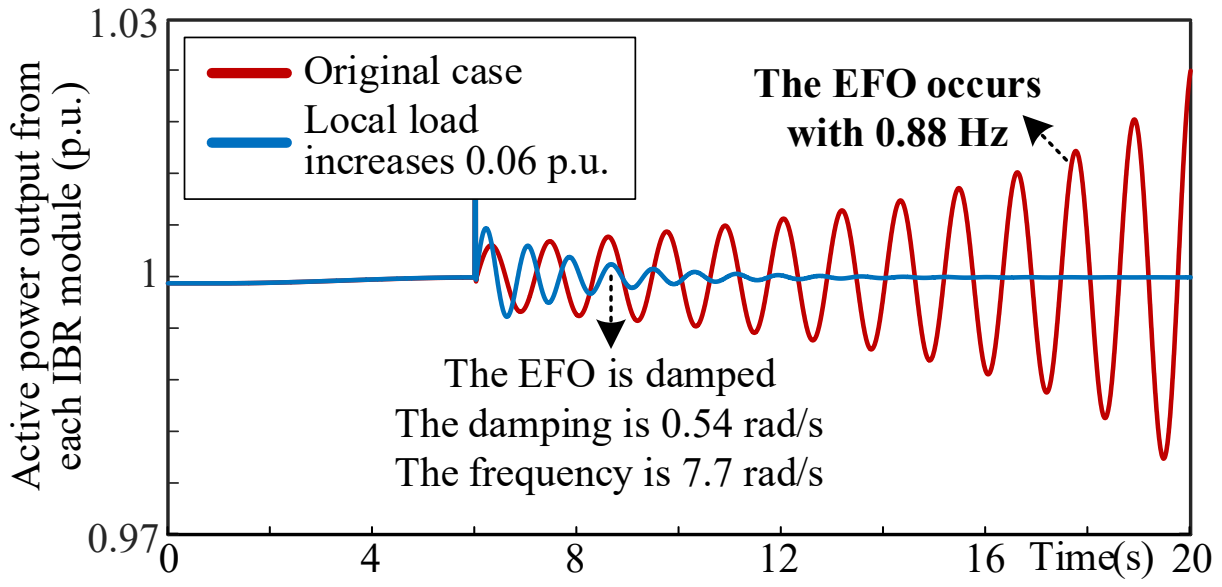

Fig. 18. Output active power of each IBR module when load power increases.

To validate the result, EMT simulations are performed with the same disturbance applied, as shown in Fig. 18. When the local active load is increased to 0.06 p.u., the active power responses remain stable after the disturbance. The simulation agrees with the modal trajectory in Fig. 17 and confirms that sufficient local active power consumption can reduce the SCRE and suppress DVC-EFOs.

## VII. CONCLUSIONS

This study reports an EFO event observed in a practical power system connected with multiple PV plants and explains why the event can be dominated by DVC loops. The key feature is that the inherent DVC dynamics remain in the subsynchronous range, whereas the closed-loop dominant mode shifts into the EFR through network-mediated interactions. This mechanism differs from traditional EFOs associated with SG dynamics, SG-like controls, PLLs, plant-level control, or manual controller retuning. The findings are not limited to PV plants. They show that IBRs with subsynchronous inherent dynamics can still induce EFOs when many similar modules interact through a network. The risk increases as the IBR scale expands or as the minimum eigenvalue of the network admittance matrix decreases. The main conclusions are as follows:

1) This study verifies, using both field evidence and analytical modeling, that GFL-IBRs can induce DVC-EFOs. The cause is the network-mediated interaction among similar IBR modules in plants, rather than the inherent control loop of a single IBR, manual controller tuning, or PLL synchronization instability. The SCRE imposes a negative synchronous contribution on the inherent DVC oscillation loop and thereby reduces the closed-loop oscillation frequency.

2) A complex vector-based analysis method is proposed to clarify the contributing factors. The SCRE depends primarily on the IBR scale, the network configuration, and local loads. A stability criterion for DVC-EFOs is derived to estimate the allowable IBR scale and the required network or load condition without retuning DVC parameters. The minimum eigenvalue of the network admittance matrix is identified as a key indicator: lower values correspond to stronger network-mediated dynamic interactions and a higher DVC-EFO risk.

3) Mitigation strategies are proposed from three perspectives: limiting the number of IBR modules, reinforcing transmission paths, and increasing local active load. For the study case, full power delivery with five IBR modules requires either an effective 3% reduction of line length or at least 0.06 p.u. local active power load under the 7% damping ratio constraint. These results indicate that IBR capacity should be coordinated with network reinforcement and local consumption. If residential demand cannot always be guaranteed, energy storage can be used to keep the combined active power consumption during high IBR output periods.

Other poorly damped oscillation modes may coexist in the EFR. For example, a slow IBR with SG-like controls or an SG with weak rotor damping may introduce its own EFR mode, while the DVC loops of large IBR plants may shift a closed-loop mode into the same frequency range through the SCRE. If these modes are close in frequency or are strongly coupled through the network, modal interaction may further reduce damping and may even contribute to inter-area oscillations. Therefore, the PLL, DVC loop, network admittance, and other poorly damped EFR modes should be examined together. Future work will extend the analysis to mixed power systems containing SGs, GFM-IBRs, and GFL-IBRs with different control bandwidths.

## APPENDIX A

TABLE V
Parameters of the study case

| Parameters | Actual values | Per-unit values |
|---|---|---|
| Number of bridge arms | 3 | / |
| Power electronic device | Three-level VSC [28] | / |
| Nominal power and frequency | [1MW, 50 Hz] | 1 p.u. |
| Nominal primary voltage (Vrms) | 35 kV | 1 p.u. |
| Nominal DC bus voltage | 750 V | 1 p.u. |
| VDC regulator gains | [1, 91.7] | [1, 0.292] |
| RPC regulator gains | [0.1, 8.3] | [0.1, 0.026] |
| Current regulator gains | [0.3, 80] | [1, 0.255] |
| AC inductance base | 3.9013 H | 1 p.u. |
| AC impedance base | 1225 Ω | 1 p.u. |
| DC impedance base | 0.5625 Ω | 1 p.u. |
| DC capacitance base | 0.005659 F | 1 p.u. |
| DC capacitance | 0.0142 F | 2.5 p.u. |
| Simulation step time | 2e-5 sec. | / |
| Control sample time | 1e-4 sec. | / |